\documentclass[12pt]{amsart}
\usepackage{amssymb}
\usepackage{amscd}
\usepackage{graphicx}

\def\({\left(}
\def\){\right)}
\def\[{\left[}
\def\]{\right]}

\newcommand{\nn}{\nonumber}
\newcommand{\bea}{\begin{eqnarray}}
\newcommand{\ena}{\end{eqnarray}}
\newcommand{\be}{\begin{eqnarray*}}
\newcommand{\en}{\end{eqnarray*}}
\newcommand{\ba}{\begin{array}}
\newcommand{\ea}{\end{array}}

\newcommand{\R}{{\mathbb R}}
\newcommand{\C}{{\mathbb C}}
\newcommand{\Z}{{\mathbb Z}} 
\newcommand{\Q}{{\mathbb Q}} 
\newcommand{\A}{{\mathcal A}} 
\newcommand{\bA}{\bar{\mathcal A}} 
\renewcommand{\S}{{\mathcal S}} 
\newcommand{\cP}{\mathcal{P}}

\newcommand{\Ac}{{\bf A^{\rm cl}}}
\newcommand{\tAc}{\tilde{\bf A}^{\rm cl}}

\newcommand{\Oc}{{\mathcal O}}
\newcommand{\Kb}{{\bf K}}
\newcommand{\Kc}{{\bf K^{\rm cl}}}
\newcommand{\Ab}{{\bf A}}
\newcommand{\Fb}{{\bf F}}
\newcommand{\Fc}{{\bf F^{\rm cl}}}
\newcommand{\Hb}{{\bf H}}
\newcommand{\Hc}{{\bf H^{\rm cl}}}

\newcommand{\Xh}{\widehat{X}}
\newcommand{\Yh}{\widehat{Y}}
\newcommand{\Oh}{\widehat{\Omega}}

\newcommand{\Rc}{\check{R}}
\newcommand{\slt}{\mathfrak{sl}_2}

\newcommand{\V}{\mathcal{V}}

\newcommand{\Tb}{{\bf T}}
\newcommand{\Rb}{{\bf R}}

\renewcommand{\Im}{\mathop{\rm Im}}
\renewcommand{\Re}{\mathop{\rm Re}}  

\newcommand{\id}{{\rm id}}
\newcommand{\tr}{{\rm tr}}

\newcommand{\Tr}{{\rm Tr}}

\newcommand{\End}{\mathop{\rm End}}

\newcommand{\la}{\lambda}

\newcommand{\al}{\alpha}

\newcommand{\ve}{\varepsilon}
\newcommand{\bs}{\mathbf{s}}
\newcommand{\om}{\omega}

\newenvironment{tenumerate}{
  \begin{enumerate}
  
  }{\end{enumerate}}
\newcommand{\bi}{\begin{tenumerate}}
\newcommand{\ei}{\end{tenumerate}}
\newcommand{\isoto}[1][]%
{{\mathop{\buildrel{\sim}\over\longrightarrow}\limits_{#1}}}

\numberwithin{equation}{section}
\newtheorem{thm}{Theorem}[section]
\newtheorem{prop}[thm]{Proposition}
\newtheorem{lem}[thm]{Lemma}

\begin{document} 
\title[Sklyanin algebra]
{Traces on the Sklyanin algebra and 
correlation functions of the eight-vertex model}
\date{\today}
\author{H.~Boos, M.~Jimbo, T.~Miwa, F.~Smirnov and Y.~Takeyama}
\address{HB: Physics Department, University of Wuppertal, D-42097,
Wuppertal, Germany\footnote{
on leave of absence from the Institute for High Energy Physics, Protvino, 
142281, Russia}}\email{boos@physik.uni-wuppertal.de}
\address{MJ: Graduate School of Mathematical Sciences, The
University of Tokyo, Tokyo 153-8914, Japan}\email{jimbomic@ms.u-tokyo.ac.jp}
\address{TM: Department of Mathematics, Graduate School of Science,
Kyoto University, Kyoto 606-8502, 
Japan}\email{tetsuji@math.kyoto-u.ac.jp}
\address{FS\footnote{Membre du CNRS}: Laboratoire de Physique Th{\'e}orique et
Hautes Energies, Universit{\'e} Pierre et Marie Curie,
Tour 16 1$^{\rm er}$ {\'e}tage, 4 Place Jussieu
75252 Paris Cedex 05, France}\email{smirnov@lpthe.jussieu.fr}
\address{YT: Graduate School of Pure and Applied Sciences, 
Tsukuba University, Tsukuba, Ibaraki 305-8571, Japan}
\email{takeyama@math.tsukuba.ac.jp}

\begin{abstract}
We propose a conjectural formula 
for correlation functions of the 
Z-invariant (inhomogeneous) eight-vertex model. 
We refer to this conjecture as {\it Ansatz}.  
It states that correlation functions 
are linear combinations of products of 
three transcendental functions, 
with theta functions and derivatives as coefficients.  
The transcendental functions 
are essentially logarithmic derivatives of the 
partition function per site. 
The coefficients are given in terms of a linear 
functional $\Tr_\la$ on the Sklyanin algebra, which interpolates 
the usual trace on finite dimensional representations. 
We establish the existence of $\Tr_\la$ and discuss the  
connection to the geometry of the classical limit. 
We also conjecture that 
the Ansatz satisfies the reduced qKZ equation.  
As a non-trivial example of the Ansatz, we present a 
new formula for 
the next-nearest neighbor correlation functions.  
\end{abstract}

\maketitle

\setcounter{section}{0}
\setcounter{equation}{0}

\section{Introduction}\label{sec:1}

Exact description of correlation functions 
and their analysis is one of the central problems of 
integrable lattice models. 
Significant progress has been made over the last decade
toward this goal. 
In the study of correlation functions,  
a basic role is played by a multiple integral representation, 
first found for the archetypical 
example of the spin $1/2$-XXZ chain \cite{JMMN,JM,KMT}.  
Subsequently it has been generalized in  
several directions, to incorporate 
an external field \cite{KMT}, unequal time \cite{KMST2},  
non-zero temperature \cite{GKS} and finite chains \cite{KMST1}.  
Earlier in the literature,  
extension to elliptic models has also been pursued. 
The free field construction used in the XXZ model was 
extended in \cite{LuP} to the SOS models, resulting in an 
integral formula for correlation functions of the ABF model.  
In \cite{LP1} an integral formula was obtained for the 
eight-vertex model by mapping the problem 
to the SOS counterpart. 
A novel free field representation 
of the eight-vertex model is being developed in \cite{Sh1,Sh2}.  

Recent studies have revealed 
another aspect of these integrals.  
Through examples at short distance, 
it has been observed in the case of the homogeneous XXX chain 
that the relevant integrals can be 
evaluated in terms of the Riemann zeta function 
at odd integers with rational coefficients \cite{BK}. 
Similar calculations have been performed for 
the XXZ chain \cite{KSTS,TKS}. 
This phenomenon was explained later through 
a duality between the qKZ equations of level $0$ and level $-4$ 
\cite{BKS1,BKS2}. 
Motivated by these works, we have established 
in our previous papers \cite{BJMST1,BJMST2} 
an algebraic representation 
(in the sense no integrals are involved) 
for general correlation functions 
of the inhomogeneous six-vertex model and its degeneration
\footnote{Correlation functions of 
the XXZ and XXX chains 
are given in the limit where all the inhomogeneity parameters 
are chosen to be the same. 
However we have not succeeded in performing this 
homogeneous limit.}. 
The aim of the present paper is to continue our study and 
examine the eight-vertex model. 

We formulate a conjectural formula for 
correlation functions (the Ansatz) 
along the same line with the six-vertex case. 
Consider the eight-vertex model where each 
column $i$ (resp. row $j$) 
carries an independent spectral parameter $t_i$ 
(resp. $0$). 
The object of our interest is the matrix 
\be
&&h_n(t_1,\cdots,t_n)
\\
&&\quad
=\frac{1}{2^n}\sum_{\alpha_1,\cdots,\alpha_n=0}^3
\ve_{a_1}\cdots\ve_{a_n}
\langle \sigma_1^{\alpha_1}\cdots\sigma_n^{\alpha_n}\rangle
\,
(\sigma^{\alpha_1}\otimes\cdots\otimes\sigma^{\alpha_n})^T
\quad\in
\End\bigl((\C^2)^{\otimes n}\bigr)
\en
where $\langle \cdots \rangle$ 
denotes the ground state average in the thermodynamic limit, 
$\sigma^0=1$,  
$\sigma^a$ ($1\le a\le 3$) are the Pauli matrices,  
and $T$ stands for the matrix transpose. 
Regard $h_n$ as a vector via the identification 
$\End\bigl((\C^2)^{\otimes n}\bigr)\simeq(\C^2)^{\otimes 2n}$, 
and let $\bs_{n}$ denote the vector corresponding to the 
identity.  
Our Ansatz is that $h_n$ can be represented in the form
\be
h_n(t_1,\cdots,t_n)=2^{-n}
\exp\left(\sum\limits _{i<j}\sum_{a=1}^3
\omega_a(t_{ij})\Xh^{(i,j)}_{a,n}(t_1,\cdots,t_n) 
\right)
\bs_{n}. 
\en
Here $\omega_a(t)$ are scalar functions 
given explicitly in terms of the partition function per site 
(see \eqref{eq:om} below). 
The matrices $\Xh^{(i,j)}_{a,n}$ are 
expressible by theta functions and derivatives.    
Leaving the details to Section \ref{sec:ansatz},  
let us comment on the latter. 

In the six-vertex case, $\Xh^{(i,j)}_{a,n}$ 
are defined in terms of a `trace' of a monodromy matrix. 
Here `trace' means the unique linear functional 
\be
\Tr_\la~:~U_q(\slt)\longrightarrow
\C[q^{\pm\la}]\oplus\la\C[q^{\pm\la}], 
\en
which for $\la\in\Z_{\ge0}$ reduces
to the usual trace on the $\la$-dimensional 
irreducible representation of $U_q(\slt)$. 
In the eight-vertex case, we need 
an analogous functional $\Tr_\la$, 
defined on the Sklyanin algebra 
and taking values in the space of entire functions involving 
$\la$, theta functions and derivatives.  
Compared with the trigonometric case, 
the existence of $\Tr_\la$ is more difficult to establish. 
We do that by considering the classical limit 
and showing that, 
for generic values of the structure constants, 
the computation of 
the trace of an arbitrary monomial can be reduced to 
that of seven basic monomials. 
We have also compared our formula for ${\rm Tr}_{\lambda}$ with the 
results by K. Fabricius and B. McCoy \cite{FM1} for $\lambda=3, 4, 5$. 
In the classical limit, the Sklyanin algebra becomes 
the algebra of regular functions 
on an algebraic surface in $\C^4$, 
which turns out to be a smoothing of 
a simple-elliptic singularity of K. Saito \cite{Sa}. 
The reduction of the trace 
is closely connected with the de Rham cohomology
of this surface (see Appendix \ref{app:existence}). 
Although we do not use 
Saito's results for our immediate purposes, 
we find this connection intriguing. 

In the trigonometric case, it was shown \cite{BJMST2}
that the functions given by the above Ansatz satisfies 
the reduced qKZ equation. 
The steps of the proof carry over 
straightforwardly to the elliptic case, 
except for one property (the Cancellation identity). 
Unfortunately 
we have not succeeded in proving this last relation. 
It remains an open question to show that 
our Ansatz in the elliptic case satisfies the
reduced qKZ equation. 

To check the validity of the Ansatz, 
we examine the simplest case $n=2$. 
In this case 
an exact answer for the homogeneous chain is obtained 
as derivatives of the ground state energy of the spin-chain Hamiltonian. 
Our formula matches with it. 
We also present an explicit formula for the correlators
with $n=3$. 
It agrees well numerically  
with the known integral formulas of \cite{LP1,LP2}. 

The plan of the paper is as follows. 
In Section 2, we introduce our notation and formulate the 
Ansatz for correlation functions. 
In Section 3, we discuss the validity of the 
Cancellation Identity and give arguments in its favor. 
Section 4 is devoted to the examples for correlators
of the nearest and the next nearest neighbor spins.  
We also discuss briefly the trigonometric limit. 
In Appendix \ref{app:existence} we prove 
the existence of the trace functional. 
As was mentioned above, the classical limit of the
Sklyanin algebra is related to an affine algebraic surface, 
and the the trace functional tends to 
an integral over a certain cycle on it. 
We explain the connection 
between this picture and K. Saito's theory. 
In Appendix \ref{app:cycles}, 
we give an explicit description of the integration cycles. 
Appendix \ref{app:lemmas} 
contains technical Lemmas about the trace. 
Finally in Appendix \ref{app:transX} 
we discuss 
the transformation properties of the matrices 
$\Xh^{(i,j)}_{a,n}$. 

\section{Ansatz for correlation functions}\label{sec:2}

In this section we introduce our notation and 
formulate the Ansatz for correlation functions 
of an inhomogeneous eight-vertex model, 
following  the scheme developed in \cite{BJMST2}. 

\subsection{$R$ matrix}\label{subsec:Rmat}

We consider an elliptic $R$ matrix 
depending on three complex parameters $t,\eta,\tau$.   
We assume 
$\Im\tau>0$ 
and $\eta\not\in\Q+\Q\tau$ 
\footnote{Later on we also assume that $\eta$ is generic}. 
We will normally regard $\eta,\tau$ as fixed constants 
and suppress them from the notation. 
Let $\theta_\alpha(t)=\theta_\alpha(t|\tau)$
($0\le \alpha\le 4$, $\theta_4(t)=\theta_0(t)$) denote 
the Jacobi elliptic theta functions 
associated with the lattice $\Z+\Z\tau$ \cite{HTF}. 
We set 
\be
&&[t]:=\frac{\theta_1(2t)}{\theta_1(2\eta)}. 
\en
The $R$ matrix is given by 
\bea
&&R(t):=\rho(t)
\frac{r(t)}{[t+\eta]}
\quad \in \End(V\otimes V), 
\label{eq:R}
\\
&&
r(t):=\frac{1}{2}\sum_{\alpha=0}^3
\frac{\theta_{\alpha+1}(2t+\eta)}{\theta_{\alpha+1}(\eta)}
\,
\sigma^{\alpha}\otimes\sigma^{\alpha}, 
\label{eq:r}
\ena
where $V=\C v_+\oplus \C v_-$. 

The matrix $r(t)=r_{12}(t)$ is the unique entire function satisfying 
\be
&&r_{12}(0)=P_{12},
\\
&&\sigma^a_1\sigma^a_2 r_{12}(t)
=r_{12}(t)\sigma^a_1\sigma^a_2
\qquad (a=1,2,3), 
\\
&&r_{12}\Bigl(t+\frac{1}{2}\Bigr)=-\sigma^1_1 r_{12}(t) \sigma^1_1,
\\
&&r_{12}\Bigl(t+\frac{\tau}{2}\Bigr)=
-\sigma^3_1 r_{12}(t) \sigma^3_1\times e^{-2\pi i(2t+\eta+\tau/2)}. 
\en
Here $P\in\End(V\otimes V)$ signifies the transposition 
$Pu\otimes v=v\otimes u$. 
As is customary, the suffix of a matrix indicates 
the tensor component on which it acts non-trivially, e.g.
$\sigma_1^\alpha=\sigma^\alpha\otimes 1$, 
$\sigma_2^\alpha=1\otimes \sigma^\alpha$.

The normalizing factor $\rho(t)$ is chosen to ensure that the partition function per site 
of the corresponding eight-vertex model equals to $1$. 
Its explicit formula depends on the regime under consideration, 
and will be given later in \eqref{eq:rho-dis},\eqref{eq:rho-ord}.
In each case it satisfies
\be
\rho(t)\rho(-t)=1,
\quad \rho(t)\rho(t-\eta)=\frac{[t]}{[\eta-t]}, 
\en

We will often write $t_{ij}=t_i-t_j$.
The basic properties of $R(t)$ are the Yang-Baxter equation
\bea
R_{12}(t_{12})R_{13}(t_{13})R_{23}(t_{23})
=
R_{23}(t_{23})R_{13}(t_{13})R_{12}(t_{12}),
\label{eq:YBE}
\ena
and 
\bea
&&
R(t)=PR(t)P, 
\label{eq:R1}
\\
&&
R(-\eta)=-2\cP^-,
\label{eq:R2}
\\
&&R_{12}(t)R_{21}(-t)=1, 
\label{eq:R3}
\\
&&
R_{12}(t)\cP^-_{23}=-R_{13}(-t-\eta)\cP^-_{23}.
\label{eq:R4}
\ena
In \eqref{eq:R2}, $\cP^-=(1-P)/2$ denotes 
the projection onto the one-dimensional 
subspace spanned by 
\be
&&s:=v_+\otimes v_--v_-\otimes v_+ \quad\in V\otimes V.  
\en
We will use also 
\be
\Rc(t)=P R(t).
\en

\subsection{Sklyanin algebra}\label{subsec:Sklyanin}

Along with the $R$ matrix, we will need 
the $L$ operator 
whose entries are generators of the Sklyanin algebra \cite{Sk1,Sk2}.

Recall that the Sklyanin algebra $\A$ is an associative unital $\C$-algebra 
defined through four generators $S_\alpha$ ($\alpha=0,1,2,3$)
and quadratic relations
\bea
&&[S_0,S_a]=iJ_{bc}(S_bS_c+S_cS_b), 
\label{eq:quad1}\\
&&[S_b,S_c]=i(S_0S_a+S_aS_0),  
\label{eq:quad2}
\ena
where $(a,b,c)$ runs over cyclic permutations of $(1,2,3)$. 
The $J_{bc}$ are the structure constants given by 
\bea
&&J_{bc}=-\frac{J_b-J_c}{J_a}=
\ve_a\frac{\theta_1(\eta)^2\theta_{a+1}(\eta)^2}
{\theta_{b+1}(\eta)^2\theta_{c+1}(\eta)^2}, 
\label{eq:J1}\\
&&J_a=\frac{\theta_{a+1}(2\eta)\theta_{a+1}}
{\theta_{a+1}(\eta)^2},
\label{eq:J2}
\ena
where
\be
\ve_2=-1,\quad \ve_\alpha=1~~(\alpha\neq 2).
\en
Here and after, theta functions without arguments stand for 
the theta zero values, 
$\theta_a=\theta_a(0)$ and $\theta_1'=\theta_1'(0)$. 

Since the defining relations are homogeneous, 
$\A$ is a $\Z_{\ge 0}$-graded algebra,  
$\A=\oplus_{n\ge 0}\A_n$,  
where the generators $S_a$ all belong to $\A_1$.  
We have also a $\Z_2\times\Z_2$-grading,  
$\A=\oplus_{(m,n)\in\Z_2\times\Z_2}\A^{(m,n)}$, 
defined by the assignment $S_{\alpha}\in\A^{\bar{\alpha}}$, where 
\bea
\bar{0}=(0,0),~~
\bar{1}=(1,0),~~
\bar{2}=(1,1),~~
\bar{3}=(0,1)\quad \in\Z_2\times\Z_2.
\label{eq:color}
\ena
To make distinction, 
the $\Z_{\ge 0}$-grading and the $\Z_2\times\Z_2$-grading 
will be referred to as 
`degree' and `color', respectively. 
Thus $S_\alpha$ has degree $1$ and color $\bar{\alpha}$. 

There are two central elements of degree $2$ and color 
$\bar{0}$,  
\bea
K_0:=\sum_{\alpha=0}^3S_\alpha^2,\quad
K_2:=\sum_{a=1}^3J_aS_a^2.   
\label{eq:Cas}
\ena
We call them Casimir elements. 

Introduce the generating function ($L$ operator)
\be
L(t):=\frac{1}{2}\sum_{\alpha=0}^3
\frac{\theta_{\alpha+1}(2t+\eta)}{\theta_{\alpha+1}(\eta)}
S_\alpha\otimes \sigma^\alpha
\quad \in \A\otimes \End(V).
\en
The defining relations \eqref{eq:quad1}, \eqref{eq:quad2} are equivalent to 
\bea
&&R_{12}(t-s)L_1(t)L_2(s)=L_2(s)L_1(t)R_{12}(t-s).
\label{eq:RLL}
\ena
{}From \eqref{eq:Cas} we have
\bea
&&L_1\Bigl(\frac{t}{2}\Bigr)
L_2\Bigl(\frac{t}{2}-\eta\Bigr)\cP^-_{12}
=-\frac{1}{4}\left(
\frac{\theta_1(\eta-t)\theta_1(\eta+t)}
{\theta_1(\eta)^2}K_0
+
\frac{\theta_1(t)^2}{\theta_1(\eta)^2}K_2\right)
\cP^-_{12}.
\label{eq:LCas}
\ena

We will be concerned with representations in series (a) of \cite{Sk2}, 
which are analogs of finite-dimensional irreducible 
representations of $\slt$. 
For each non-negative integer $k$, 
let $\V^{(k)}$ denote the vector space of entire functions 
$f(u)$ with the properties
\be
f(u+1)=f(u)=f(-u),\quad f(u+\tau)=e^{-2\pi i k(2u+\tau)}f(u).
\en
We have $\dim \V^{(k)}=k+1$. 
The following formula defines a 
representation $\pi^{(k)}:\A\rightarrow\End(\V^{(k)})$ \cite{Sk2}:
\footnote{We have modified eq.(6) of \cite{Sk2} by a factor 
$\theta_1(2\eta)$.}
\bea
&&\left(\pi^{(k)}(S_\alpha)f\right)(u)
\label{eq:rep(a)}
\\
&&\quad =
\frac{\sqrt{\ve_\alpha}\theta_{\alpha+1}(\eta)}
{\theta_1(2\eta)\theta_1(2u)}
\left(\theta_{\alpha+1}(2u-k\eta)e^{\eta\partial_u}
-\theta_{\alpha+1}(-2u-k\eta)e^{-\eta\partial_u}\right)f(u).
\nn
\ena
Here $\sqrt{\ve_2}=i$, $\sqrt{\ve_\alpha}=1$ ($\alpha\neq 2$), and 
$(e^{\pm\eta\partial_u}f)(u)=f(u\pm\eta)$. 
In particular, if $k=1$, then in an appropriate basis we have 
\be
\pi^{(1)}(S_\alpha)=\sigma^\alpha,
\quad
\left(\pi^{(1)}\otimes\id\right)L(t)=r(t).
\en
On $\V^{(k)}$, the Casimir elements $K_0,K_2$ act as scalars
$K_0(k+1),K_2(k+1)$ respectively, where
\bea
K_0(\la)=
4\frac{\theta_1\bigl(\la\eta\bigr)^2}{\theta_1(2\eta)^2},
\quad
K_2(\la)=
4\frac{\theta_1\bigl(\la\eta+\eta\bigr)
\theta_1(\la\eta-\eta)}{\theta_1(2\eta)^2}.
\label{eq:Casval}
\ena

\subsection{The functional $\Tr_\la$}\label{subsec:trace}

In order to formulate the Ansatz, 
we need to consider the trace $\tr_{\V^{(k)}}\pi^{(k)}(A)$
of an element $A\in\A$ 
as a function of the dimension $k+1$.   
The precise meaning is as follows. 

For each $A\in\A$ one can assign 
a unique entire function $\Tr_\la A$ in $\la$ 
with the following properties:
\begin{enumerate}
\item $\Tr_\la A\bigl|_{\la=k+1}=\tr_{\V^{(k)}}\pi^{(k)}(A)$ 
holds for all $k\in\Z_{\ge 0}$,  
\item If $A\in\A_n$, $\Tr_\la A$ has the functional form
\bea
\Tr_{\frac{t}{\eta}} A=
\theta_1(t)^n\times 
\begin{cases}
g_{A,0}(t) & (\mbox{ $n$: odd }),\\
g_{A,1}(t)-\frac{t}{\eta} g_{A,2}(t) & (\mbox{ $n$: even }),\\
\end{cases}
\label{eq:trdec}
\ena
where 
$g_{A,0}(t)$, $g_{A,2}(t)$ and  
$g_{A,3}(t):=g_{A,1}(t+\tau)-g_{A,1}(t)$ 
are elliptic functions with periods $1,\tau$. 
In addition, $g_{A,1}(t+1)=g_{A,1}(t)$. 
\end{enumerate}

For example, 
\bea
&&\Tr_\la 1=\la,
\label{eq:trSa0}
\\
&&\Tr_\la S_\alpha= 2\delta_{\alpha0}
\frac{\theta_1(\la\eta)}{\theta_1(2\eta)},
\label{eq:trSa1}
\\
&&\Tr_\la S_\alpha^2
=\frac{2}{\theta_1'\theta_1(2\eta)^3}
\left(F_{\alpha 1}(\la\eta)-\la F_{\alpha 2}(\la\eta)\right),
\label{eq:trSa2}
\ena
where 
\bea
&&F_{\alpha 1}(t)=\ve_\alpha \theta_{\alpha+1}(\eta)^2
\frac{\partial}{\partial t}\left(
\theta_{\alpha+1}(t+\eta)\theta_{\alpha+1}(t-\eta)\right),
\label{eq:fa1}
\\
&&F_{\alpha2}(t)=\ve_\alpha\theta_{\alpha+1}(\eta)^2
\frac{\partial}{\partial \eta}\left(
\theta_{\alpha+1}(t+\eta)\theta_{\alpha+1}(t-\eta)\right).
\label{eq:fa2}
\ena
For reference we set 
\bea
&&F_{\alpha3}(t)=\ve_\alpha\theta_{\alpha+1}(\eta)^2
\theta_{\alpha+1}(t+\eta)\theta_{\alpha+1}(t-\eta).
\label{eq:fa3}
\ena
$\Tr_\la$ satisfies also 
\bea
&&
\Tr_\la(AB)=\Tr_\la(BA), 
\label{eq:cycl}\\
&&
\Tr_\lambda\left(K_i A\right)
=K_i(\la)\Tr_\la(A)\quad (i=0,2),
\label{eq:TrCas}
\\
&&\Tr_\la A=0
\qquad (A\in\A^{(m,n)}, (m,n)\neq (0,0)).
\label{eq:color0}
\ena
The derivation of \eqref{eq:color0} as well as 
\eqref{eq:trSa1}, \eqref{eq:trSa2} is sketched in  
Appendix \ref{app:lemmas}. 
In Appendix \ref{app:existence} we show that, 
for generic $\eta$, 
any element $A\in\A/\sum_{\alpha=0}^3[S_\alpha,\A]$ can be written as 
a $\C[K_0,K_2]$-linear combination of 
seven monomials: $1,S_0, S_1,S_2,S_3,S_0^2,S_3^2$. 
Hence $\Tr_\la A$ is completely determined 
by the property \eqref{eq:cycl}--\eqref{eq:color0}
along with \eqref{eq:trSa0}--\eqref{eq:trSa2}.
However an effective algorithm for the reduction
is not known to us.  
The situation is in sharp contrast to the trigonometric case,  
where a simple recursive procedure
for calculating $\Tr_\la$ is available (see \cite{BJMST2}). 
It would be useful if one can 
find a more direct expression for the trace 
using $Q$-operators, 
as is done for the XXZ model in \cite{Korff2}.  

In this connection, notice that by the above rule, 
$\Tr_\la A$ has a simpler structure 
for elements of odd degree than those of even degree, 
since in the former case it is a polynomial of 
$\theta_1(t+c)$ ($c\in\C$) without involving derivatives. 
The subtle difference between finite spin chains with odd 
length and those with even length
has been noticed in the context of $Q$-operators 
\cite{PStr, FM2, Korff}. 

\subsection{The Ansatz}\label{sec:ansatz}

Consider an inhomogeneous eight vertex model, where each 
column $i$ (resp. row $j$) 
carries a spectral parameter $t_i$ (resp. $0$).   
The Boltzmann weights are given by the entries 
$R_{\ve_1,\ve_2}^{\ve'_1,\ve'_2}(t_i)$ 
of the $R$ matrix \eqref{eq:R}.   
We choose the normalizing factor $\rho(t)$ in \eqref{eq:R} in 
accordance with the two regimes 
\begin{enumerate}
\item $\eta,t\in i\R$, $-i\eta>0$  (disordered regime)
\item $\eta,t\in \R$, $\eta<0$ (ordered regime)
\end{enumerate}
In the disordered regime,  $\rho(t)=\rho^{dis}(t)$ is given by \cite{Baxbk}
\bea
&&\rho^{dis}(t):=e^{-2\pi it}
\times
\frac{\gamma(2\eta-2t)}
{\gamma(2\eta+2t)}
\frac{\gamma(4\eta+2t)}
{\gamma(4\eta-2t)},
\label{eq:rho-dis}
\\
&&\gamma(u)=\Gamma(u,4\eta,\tau),
\nn
\ena
where 
\be
\Gamma(u,\sigma, \tau):=
\prod_{j,k=0}^\infty
\frac{1-e^{2\pi i((j+1)\tau+(k+1)\sigma-u)}}
{1-e^{2\pi i(j\tau+k\sigma+u)}}
\en
is the elliptic gamma function \cite{Ruij}. 
In the ordered regime, the formula for
$\rho(t)=\rho^{ord}(t)$ is changed to 
\bea
\rho^{ord}(t)=e^{-4\pi i \eta t/\tau}\rho'(t'),
\label{eq:rho-ord}
\ena
where $\rho'(t')$ is given by the right hand side of \eqref{eq:rho-dis}
with $t,\eta,\tau$ being replaced by
\bea
t'=\frac{t}{\tau},\quad
\eta'=\frac{\eta}{\tau},\quad
\tau'=-\frac{1}{\tau},
\label{eq:modular}
\ena
respectively. 

By correlation functions 
we mean the ground state averages 
$\langle \sigma_1^{\alpha_1}\cdots\sigma_n^{\alpha_n}\rangle$  
of a product of spin operators on 
consecutive columns $1,\ldots,n$ on a same row 
of the lattice. 
The thermodynamic limit is assumed. 
We arrange them into a matrix 
\be
&&
h_n(t_1,\cdots,t_n)
\\
&&\quad
=\frac{1}{2^n}\sum_{\alpha_1,\cdots,\alpha_n=0}^3
\langle \sigma_1^{\alpha_1}\cdots\sigma_n^{\alpha_n}\rangle
\,
(\sigma^{\alpha_1})^T\otimes\cdots\otimes(\sigma^{\alpha_n})^T
\quad\in\End(V^{\otimes n}),   
\en
where $(\sigma_i^\alpha)^T=\varepsilon_\alpha\sigma_i^\alpha$ 
stands for the transposed matrix.  
Because of the `Z-invariance' \cite{BaxZ}, 
it does not depend on $t_i$ with $i<1$ or $i>n$. 
When $t_1=\cdots=t_n$, each  
$\langle \sigma_1^{\alpha_1}\cdots\sigma_n^{\alpha_n}\rangle$ is
a correlation function of the infinite XYZ spin chain 
\be
H_{XYZ}=\sum_{j=-\infty}^\infty
\left(I^1 \sigma^1_j\sigma^1_{j+1} +
I^2 \sigma^2_j\sigma^2_{j+1} +
I^3 \sigma^3_j\sigma^3_{j+1} \right)
\en 
at zero temperature (the coefficients $I^a$
will be given below in \eqref{eq:Ia}).  
The $h_n$ may be viewed as 
the density matrix of a finite sub-system of length $n$,  
regarding the rest of the spins as an environment. 

{}From now on we fix $n$, and write $\bar j=2n-j+1$. 
We regard $h_n$ as a $2^{2n}$-dimensional vector 
through the identification 
$\End(V^{\otimes n})\simeq V^{\otimes 2n}$ 
given by 
\bea
E_{\ve_1,\bar{\ve}_1}\otimes \cdots\otimes E_{\ve_n,\bar{\ve}_n}
\mapsto 
\Bigl(\prod_{j=1}^n \bar{\ve}_j\Bigr)\,
v_{\ve_1}\otimes\cdots\otimes v_{\ve_n}
\otimes v_{-\bar{\ve}_n}\otimes\cdots\otimes
v_{-\bar{\ve}_1},
\label{eq:matrixform}
\ena
where $E_{\ve,\ve'}
=\left(\delta_{\ve,\alpha}\delta_{\ve',\beta}\right)_{\alpha,\beta=\pm}$.

Let us explain the constituents 
which enter the Ansatz. 

First, we define three functions 
in terms of the factor $\rho(t)$ 
(given in \eqref{eq:rho-dis} or \eqref{eq:rho-ord}) by 
\bea
\om_1(t):=
\frac{\partial }{\partial t}\log\varphi(t),
\quad
\om_2(t):=
\frac{\partial }{\partial \eta}\log\varphi(t),
\quad
\om_3(t):=\frac{\partial }{\partial \tau}\log\varphi(t),
\label{eq:om}
\ena
where we have set 
\be
\varphi(t):=\rho(t)^4\cdot \frac{\theta_1(2\eta-2t)}{\theta_1(2\eta+2t)}.
\en
They are a meromorphic solution of the system of 
difference equations
\be
&&
\omega_1(t-\eta)+\omega_1(t)=q_1(t),
\\
&&
\omega_2(t-\eta)+\omega_2(t)-\omega_1(t-\eta)=q_2(t),
\\
&&
\omega_3(t-\eta)+\omega_3(t)=q_3(t),
\en
where
\bea
q_1(t):=\frac{\partial }{\partial t}\log\psi(t),
\quad
q_2(t):=\frac{\partial }{\partial \eta}\log\psi(t),
\quad
q_3(t):=\frac{\partial }{\partial \tau}\log\psi(t),
\label{eq:q}
\ena
and 
\be
\psi(t):=\frac{\theta_1(2t)^3\theta_1(2t-4\eta)}
{\theta_1(2t-2\eta)^3\theta_1(2t+2\eta)}.
\en

The next ingredient are the matrices $\Xh_{a,n}^{(i,j)}$
($a=1,2,3$, $1\le i\neq j\le n$).  
Consider a `transfer matrix' 
\bea
&&\Xh_n(t_1,\cdots,t_n)
\label{eq:X12}
\\
&&\quad:=
\frac1{[t_{1,2}]\prod_{p=3}^n[t_{1,p}][t_{2,p}]}
\Tr_{t_{1,2}/\eta}
\Bigr(T^{[1]}_n\bigl(\frac{t_1+t_2}{2};t_1,\cdots,t_n\bigr)
\Bigl)
P_{12}\cP^-_{1\bar1}\cP^-_{2\bar2}.   
\nn
\ena
We used the functional 
$\Tr_\la$ introduced in the previous section, and  
\be
&&T^{[1]}_n(t;t_1,\ldots,t_n)
\\
&&\quad :=
L_{\bar2}(t-t_2-\eta)\cdots L_{\bar n}(t-t_n-\eta)
L_n(t-t_n)\cdots L_2(t-t_2). 
\en
Notice the presence of the permutation $P_{12}$ and the 
projectors $\cP^-_{1\bar1}\cP^-_{2\bar2}$ in \eqref{eq:X12}.

For $i<j$, we define 
\bea
\Xh_n^{(i,j)}(t_1,\cdots,t_n)
&=&\Xh_n^{(j,i)}(t_1,\cdots,t_n)
\nn
\\
&:=&\mathbb{R}_n^{(i,j)}(t_1,\cdots,t_n)
\Xh_n(t_i,t_j,t_1,
\cdots,\widehat{t_i},\cdots,\widehat{t_j},\cdots,t_n) 
\label{eq:Xij}\\
&\times&
\mathbb{R}_n^{(i,j)}(t_1,\cdots,t_n)^{-1}.
\nn
\ena
Here $\mathbb{R}_n^{(i,j)}$ stands for the product of $R$ matrices
\begin{eqnarray}
&&\mathbb{R}_n^{(i,j)}(t_1,\cdots,t_n)
\label{eq:complicated}
\\
&&:=
\Rc_{i,i-1}(t_{i,i-1})\cdots \Rc_{2,1}(t_{i,1}) 
\nn\\ 
&& {}\times 
\Rc_{j,j-1}(t_{j,j-1})\cdots 
\Rc_{i+2,i+1}(t_{j,i+1})\cdot
\Rc_{i+1,i}(t_{j,i-1})\cdot
\cdots \Rc_{3,2}(t_{j,1})
\nn\\
&&{}\times
\Rc_{\overline{i-1},\bar i}(t_{i-1,i})\cdots
\Rc_{\bar 1\bar 2}(t_{1,i}) 
\nn\\ 
&& {}\times 
\Rc_{\overline{j-1},\bar{j}}(t_{j-1,j})\cdots 
\Rc_{\overline{i+1},\overline{i+2}}(t_{i+1,j})
\cdot 
\Rc_{\overline{i},\overline{i+1}}(t_{i-1,j})
\cdots \Rc_{\bar{2},\bar{3}}(t_{1,j}).
\nn
\end{eqnarray}
Finally, for all $i\not=j$, $\Xh_{a,n}^{(ij)}$ are defined by 
\bea
&&
c\Xh^{(i,j)}_{1,n}(t_1,\cdots,t_n)
:=
\Xh^{(i,j)}_n(t_1,\cdots ,t_n)-
t_{ij}\Delta _1^{(i)}\Xh^{(i,j)}_n(t_1,\cdots ,t_n),
\label{eq:Xn1}\\
&&
c\Xh^{(i,j)}_{2,n}(t_1,\cdots,t_n)
=-\eta\Delta _1^{(i)}\Xh^{(i,j)}_n(t_1,\cdots ,t_n),
\label{eq:Xn2}\\
&&
c\Xh^{(i,j)}_{3,n}(t_1,\cdots,t_n)
:=\Delta _\tau^{(i)}
\Xh^{(i,j)}_n(t_1,\cdots ,t_n)
-\tau\Delta _1^{(i)}\Xh^{(i,j)}_n(t_1,\cdots ,t_n),
\label{eq:Xn3}
\ena
where $c=-2\theta_1'/\theta_1(2\eta)$ and  
\be
\Delta _a^{(i)}f(\cdots,t_i, \cdots)=
f(\cdots,t_i+a, \cdots)-f(\cdots ,t_i ,\cdots). 
\en
As we show in Appendix \ref{app:transX}, 
the $\Xh_{a,n}^{(ij)}(t_1,\cdots,t_n)$ 
are doubly periodic in $t_k$ with periods $1, \tau$.  
The only exception is the case $a=1$, $k=i$ or $j$ and 
with respect to the shift by $\tau$, 
where the transformation law becomes
\be
&&\Delta ^{(i)}_\tau\Xh^{(i,j)}_{1,n}(t_1,\cdots,t_n)
=\Delta ^{(j)}_{-\tau}
\Xh^{(i,j)}_{1,n}(t_1,\cdots,t_n)
=\Xh_{3,n}^{(i,j)}(t_1,\cdots,t_n).
\en
Conversely we have 
\be
\Xh^{(ij)}_n(t_1,\cdots,t_n)
&=&
c\left(
\Xh_{1,n}^{(i,j)}(t_1,\cdots,t_n)
-\frac{t_{ij}}{\eta}\Xh_{2,n}^{(i,j)}(t_1,\cdots,t_n)\right).
\en
We compute the trace in the formula (\ref{eq:Xij})
by using the formulas (\ref{eq:Casval}) and (\ref{eq:trSa2}).
The separation of $\Xh_n^{(i,j)}$
into two parts $\Xh_{1,n}^{(i,j)}$ and $\Xh_{2,n}^{(i,j)}$
comes from that of $\Tr_\la S_\alpha^2$ into 
$F_{\alpha 1}(\la\eta)$ and $F_{\alpha 2}(\la\eta)$.
Note that $\Xh^{(j,i)}_{1,n}=\Xh^{(i,j)}_{1,n}$ and 
$\Xh^{(j,i)}_{a,n}=-\Xh^{(i,j)}_{a,n}$ for $a=2,3$.

We are now in a position to state our conjecture. 
Let 
\be
&&{\bf s}_n:=\prod_{p=1}^ns_{p\bar p}
\en
be the vector corresponding to the identity by the map
\eqref{eq:matrixform}.
\medskip

{\bf Conjecture.}\quad 
Correlation functions of the inhomogeneous eight-vertex model 
are given by the formula 
\bea
&&
h_n(t_1,\cdots,t_n)=2^{-n}
\exp\left(\sum\limits _{i<j}\sum_{a=1}^3
\omega_a(t_{ij})\Xh^{(i,j)}_{a,n}(t_1,\cdots,t_n) \right)
\bs_{n}, 
\label{eq:main}
\ena
where $\omega_a(t)$ and $\Xh^{(i,j)}_{a,n}$ are defined 
respectively by
\eqref{eq:om}
and \eqref{eq:X12}--\eqref{eq:Xn3}. 
\qed 

\section{Reduced qKZ equation}

The $h_n$ is known to satisfy the following set of equations 
\cite{JMN}:  
\bea
&&
h_n(\cdots,t_{j+1},t_j,\cdots)
\label{eq:Rsymm}
\\
&&{}=
\Rc_{j,j+1}(t_{j,j+1})
\Rc_{\overline{j+1},\bar{j}}(t_{j+1,j})
h_n(\cdots,t_j,t_{j+1},\cdots)
\quad (1\le j\le n-1),\nn
\\
&&h_n(\cdots, t_j-\eta,\cdots)
=A^{(j)}_n(t_1,\cdots,t_n)h_n(\cdots,t_j,\cdots),
\label{eq:rqKZ}
\\
&&
\cP^-_{1,\bar{1}}\cdot 
h_n(t_1,\cdots,t_n)_{1,\ldots,n,\bar{n},\ldots,\bar{1}}
=\frac{1}{2}s_{1\bar1}
h_{n-1}(t_2,\cdots,t_n)_{2,\ldots,n,\bar{n},\ldots,\bar{2}}.
\label{eq:n_to_n-1}
\ena
Here 
\bea
\quad&& A^{(j)}_n(t_1,\cdots,t_n)
\label{eq:Aj}\\
&&
=(-1)^n
R_{j,j-1}(t_{j,j-1}-\eta)\cdots R_{j,1}(t_{j,1}-\eta)
R_{\bar j,\overline{j+1}}(t_{j,j+1}-\eta)\cdots 
R_{\bar j,\bar n}(t_{j,n}-\eta)
\nn\\
&&\quad\times
P_{j,\bar j}
R_{j,n}(t_{j,n})\cdots R_{j,j+1}(t_{j,j+1})
R_{\bar j,\bar1}(t_{j,1})\cdots 
R_{\bar j,\overline{j-1}}(t_{j,j-1})\,. 
\nn
\ena
In this section,  assuming a conjectural identity,  
we explain that these relations are valid also for the Ansatz. 

\subsection{Properties of $\Oh^{(i,j)}_n$}

Consider the expression 
\be
\widehat{\Omega}^{(i,j)}_n(t_1, \cdots , t_n)=
\sum_{a=1}^3
\omega_a(t_{ij})\Xh^{(i,j)}_{a,n}(t_1,\cdots,t_n), 
\en
which enters the Ansatz \eqref{eq:main}. 
In \cite{BJMST2} for the XXZ model, 
the following relations are derived. 

\begin{description}
\item[Exchange relation]
\bea
&&
\Rc_{k,k+1}(t_{k,k+1})\Rc_{\overline{k+1},\overline{k}}(t_{k+1,k})
\Oh_n^{(i,j)}(\cdots,t_{k},t_{k+1},\cdots)
\label{eq:exchange}
\\&&=
\Oh_n^{(\pi_k(i),\pi_k(j))}(\cdots,t_{k+1},t_k,\cdots)
\Rc_{k,k+1}(t_{k,k+1})\Rc_{\overline{k+1},\overline{k}}(t_{k+1,k}),
\nn
\ena
Here $\pi_k$ signifies the transposition $(k,k+1)$. 
\end{description}
\begin{description}
\item[Difference equations] 
\bea
&&
\Oh_{n}^{(i,j)}(t_1,\cdots,t_k-\eta,\cdots,t_n)
\label{eq:Oh-diff}\\
&&{}\quad =A^{(k)}_n(t_1,\cdots,t_n) \Oh_n^{(i,j)}(t_1,\cdots,t_n)
A^{(k)}_{n}(t_1,\cdots,t_n)^{-1}
\quad (k\neq i,j), 
\nn\\
&&\Oh^{(i,j)}_n(t_1,\cdots,t_i-\eta,\cdots,t_n)\, 
{\bf s}_{n}
\nn
\\
&&\quad 
=A^{(i)}_n(t_1,\cdots,t_n)
\left(\Oh^{(i,j)}_n(t_1,\cdots,t_n)+
\Yh^{(i,j)}_n(t_1,\cdots,t_n)\right){\bf s}_{n}.\nn
\ena
In the last line, we have set
\be
\Yh^{(i,j)}_n(t_1,\cdots,t_n):=
\sum_{a=1}^3q_a(t_{ij})\,\Xh^{(i,j)}_{a,n}(t_1,\cdots,t_n), 
\en
where $q_a(t)$ are given by 
\eqref{eq:q}.

\end{description}
\begin{description}
\item[Recurrence relation]
\begin{eqnarray}
&& 
\mathcal{P}_{1, \bar{1}}^{-}\,
\widehat{\Omega}_{n}^{(i,j)}(t_{1}, \cdots , t_{n}) \\ 
&& {}=
\left\{ 
\begin{array}{ll}
0 & (1=i<j\le n), \\
\widehat{\Omega}_{n-1}^{(i-1, j-1)}
(t_{2}, \cdots , t_{n})_{2, \ldots , n, \bar{n}, \ldots , \bar{2}} 
\, \mathcal{P}_{1, \bar{1}}^{-} & 
(2\le i<j \le n).
\end{array}\right.
\nonumber  
\end{eqnarray}
\end{description}

\begin{description}
\item[Commutativity]
For distinct indices $i,j,k,l$, 
\bea
&&
\Oh_n^{(i,j)}(t_1,\cdots,t_n)\Oh_n^{(k,l)}(t_1,\cdots,t_n)
=
\Oh_n^{(k,l)}(t_1,\cdots,t_n)\Oh_n^{(i,j)}(t_1,\cdots,t_n).
\label{eq:commutativity}
\ena
\end{description}
\begin{description}
\item[Nilpotency]
\begin{align}
&
\Oh_n^{(i,j)}(t_1,\cdots,t_n)\Oh_n^{(k,l)}(t_1,\cdots,t_n)
=0
\quad \mbox{if $\quad$ $\{i,j\}\cap\{k,l\}\neq\emptyset$}. 
\label{eq:nilpot}
\end{align}

\end{description}
The proof of these relations given in \cite{BJMST2} 
are based only on the 
properties \eqref{eq:YBE}, \eqref{eq:R1}--\eqref{eq:R4} 
of the $R$ matrix and 
\eqref{eq:RLL}, \eqref{eq:LCas}
of the $L$ operator.
Hence they carry over to the elliptic case as well.

As is shown in \cite{BJMST2}, Proposition 4.1, 
the equations \eqref{eq:exchange}--\eqref{eq:nilpot}
guarantee the validity of the fundamental properties 
\eqref{eq:Rsymm}, \eqref{eq:rqKZ}, \eqref{eq:n_to_n-1} 
for the Ansatz, 
provided one additional identity holds:
\begin{description}
\item[Cancellation identity]
\begin{eqnarray}
\(\sum_{j=2}^n
\Yh^{(1,j)}_n(t_1,\cdots,t_n)+
\left(A^{(1)}_n(t_1,\cdots,t_n)^{-1}-1\right)\)
\bs_n=0. 
\label{eq:X-last}
\end{eqnarray} 
\end{description}

So far we have not been able to prove the 
cancellation identity. 
In the next subsection, 
we suggest a possible approach toward its proof.

\subsection{Cancellation identity}

Set 
\bea
&&Q^{(i)}_n(t_1,\cdots,t_n)
\label{eq:;QY}
\\
&&\quad=
\(\sum_{j=2}^n
\Yh^{(1,j)}_n(t_1,\cdots,t_n)+
\left(A^{(1)}_n(t_1,\cdots,t_n)^{-1}-1\right)\)
\bs_n.
\nn
\ena
We regard it as a matrix via the        
isomorphism \eqref{eq:matrixform}. 

Besides the obvious translation invariance, 
$Q_n=Q_n^{(1)}$ has the following properties. 
\bea
&&
\mbox{$\prod_{j=2}^n\theta_1(2t_{1,j})
\cdot Q_n(t_1,\cdots,t_n)$ is entire}, 
\label{eq:Qhol}
\\
&&
Q_n(\cdots,t_j+\frac{1}{2},\cdots)=
\sigma^1_jQ_n(\cdots,t_j,\cdots)\sigma^1_j,
\quad (1\le j\le n),
\label{eq:Qper1}
\\
&&
Q_n(\cdots,t_j+\frac{\tau}{2},\cdots)=
\sigma^3_jQ_n(\cdots,t_j,\cdots)\sigma^3_j,
\quad (1\le j\le n),
\label{eq:Qper2}
\\
&&
\check{R}_{j,j+1}(t_{j,j+1})Q_n(\cdots,t_j,t_{j+1},\cdots)
\label{eq:Qexchg}
\\
&&\quad
=Q_n(\cdots,t_{j+1},t_{j},\cdots)\check{R}_{j,j+1}(t_{j,j+1}),
\quad (2\le j\le n-1),
\nn\\
&&
Q_n(t_1,\cdots,t_{n-1},t_{n})
\cP^-_{n-1,n}\Bigl|_{t_{n-1}=t_{n}+\eta}=
Q_{n-2}(t_1,\cdots,t_{n-2})\cP^-_{n-1,n},
\label{eq:Qred1}
\\
&&{\rm tr}_1 Q_n(t_1,\cdots,t_n)=0,
\label{eq:Qred2}
\\
&&{\rm tr}_n Q_n(t_1,\cdots,t_n)
=Q_{n-1}(t_1,\cdots,t_{n-1}).
\label{eq:Qred3}
\ena
These relations are verified in  a way similar to 
those in \cite{BJMST2}. 
The derivation of \eqref{eq:Qper1}--\eqref{eq:Qper2} 
rests on the transformation laws 
of the $\Xh^{(i,j)}_{a,n}$, which we discuss in Appendix 
\ref{app:transX}. 

{}From the properties \eqref{eq:Qhol}, \eqref{eq:Qper1}, \eqref{eq:Qper2},  
$Q_n$ can be written as 
\bea
&&\prod_{j=2}^n \theta_1(2t_{1j})
\times Q_n(t_1,\cdots,t_n)
\label{eq:Qform}
\\
&&\quad=
\sum_{\alpha_1,\cdots,\alpha_n=0}^3
\kappa_{\alpha_n,\cdots,\alpha_1}
\prod_{j=1}^n\frac{\theta_{\alpha_j+1}(2t_{1j})}
{\theta_{\alpha_j+1}(\eta)}
\sigma^{\alpha_1}_1\cdots\sigma^{\alpha_n}_n,
\nn
\ena
with some $\kappa_{\alpha_n,\cdots,\alpha_1}\in\C$. 
Terms with $\alpha_1=0$ are actually absent in the sum, 
in accordance with \eqref{eq:Qred2}.  
For convenience we set $\kappa_{\alpha_n,\cdots,\alpha_2,0}=0$. 
Note that \eqref{eq:Qper1}, \eqref{eq:Qper2} 
and the translation invariance imply
\bea
\kappa_{\alpha_n,\cdots,\alpha_1}=0
\quad \mbox{ unless $\sum_{j=1}^n\bar{\alpha}_j=(0,0)$}. 
\label{eq:Qfirst}
\ena

By induction, assume $Q_m=0$ for $m<n$. 
We are going to argue that $Q_n$ is then 
determined up to a multiplicative constant 
(see Lemma \ref{lem:kappa} below). 

By \eqref{eq:Qred1}, 
the induction hypothesis and 
\eqref{eq:Qexchg},  we have
\bea
Q_n(\cdots,t_{j},t_{j+1},\cdots)
\cP^-_{j,j+1}\Bigl|_{t_j=t_{j+1}+\eta}=0
\quad (2\le j\le n-1). 
\label{eq:Qred4}
\ena
By \eqref{eq:Qred3} we may also assume
\bea
\kappa_{\alpha_n,\cdots,\alpha_1}=0
\quad \mbox{ unless }\alpha_n\neq0.
\label{eq:Qlast}
\ena

Quite generally, consider a matrix of the form 
\be
U_{1,2}(u,v)=\sum_{\alpha,\beta=0}^3
\kappa_{\beta\alpha}
\frac{\theta_{\alpha+1}(2u)}{\theta_{\alpha+1}(\eta)}
\frac{\theta_{\beta+1}(2v)}{\theta_{\beta+1}(\eta)}
\sigma^{\alpha}_1\sigma^{\beta}_2. 
\en
Then the relations 
\bea
&&\check{R}_{12}(u-v)U_{1,2}(u,v)
=U_{1,2}(v,u)\check{R}_{12}(u-v),
\label{eq:U1}
\\
&&U_{1,2}(u+\eta,u)\cP^-_{1,2}=0,
\label{eq:U2}
\ena
are equivalent to the following relations for the 
coefficients $\kappa_{ba}$:
\be
&&\kappa_{a,0}-\kappa_{0,a}=iJ_{bc}(\kappa_{c,b}+\kappa_{b,c}),
\\
&&\kappa_{b,a}-\kappa_{a,b}=i(\kappa_{c,0}+\kappa_{0,c}),
\\
&&\sum_{\alpha=0}^3\kappa_{\alpha,\alpha}=0,
\\
&&\sum_{a=1}^3J_a\kappa_{a,a}=0.
\en
Here $a,b,c$ are cyclic permutations of $1,2,3$, and 
$J_a$, $J_{bc}$ are as in \eqref{eq:J1}, \eqref{eq:J2}.
The above relations have the same form 
as those derived from the quadratic 
relations \eqref{eq:quad1},\eqref{eq:quad2} 
and from the Casimir elements \eqref{eq:Cas}, respectively. 
Consider the quotient 
$\bA$
of the Sklyanin algebra modulo the relations that 
the Casimir elements %
are 
zero. 
This is a graded algebra, 
$$
\bA
=\bigoplus\limits _{n=0}^{\infty}
\bA_n.
$$

{}From the above considerations one easily concludes that 
there exist three linear functionals $\kappa _a$ ($a=1,2,3$) 
on $\bA_{n-1}$
such that
\begin{align}
&\kappa _{\alpha_n,\cdots ,\alpha_2,a}=
\kappa _{a}\(S_{\alpha_n}\cdots S_{\alpha_2}\)\nn
\end{align}
which  satisfy the additional condition
\begin{align}
\kappa _a\(S_0A\)=0. 
\nn
\end{align}

In the Sklyanin algebra with generic parameter $\eta$, 
any monomial $S_{\alpha_2}\cdots S_{\alpha_n}$ can be reduced 
to a linear combination of ordered monomials  
$S_0^{\nu_0}S_3^{\nu_3}S_1^{\nu_2}S_2^{\nu_1}$
with $\nu_1,\nu_2\in\{0,1\}$,  
by using the quadratic relations and 
Casimir elements (PBW basis) \cite{FO,O}. 
Together with  \eqref{eq:Qfirst} this means
that each functional $\kappa _a$ is 
defined by one constant, that is, 
\begin{align}
&\kappa _1(S_3^{n-2}S_1),\ \kappa _2(S_3^{n-2}S_2),\ \kappa _3(S_3^{n-1})
\qquad\quad  \ \text{for $n$ even},
\label{const} \\
&\kappa _1(S_3^{n-2}S_2),\ \kappa _2(S_3^{n-2}S_1),\ \kappa _3(S_3^{n-3}S_1S_2)
\qquad \text{for $n$ odd}. 
\nonumber 
\end{align}
There remain {\it three} coefficients. 
In order to finish the proof of the Cancellation Identity, 
it remains to show that these coefficients vanish. 

In addition to \eqref{eq:Qexchg}, we have also the relation 
\be
\check{R}_{12}(\la_{12})Q_n(t_1,t_{2},\cdots)
\check{R}_{12}(\la_{12})^{-1}
=Q^{(2)}_n(t_{2},t_{1},\cdots). 
\en
The poles of the $R$ matrix in the left hand side 
are not the poles of $Q^{(2)}_n$. 
This entails the relation 
\be
\cP^{-}_{12}Q_n(t_1,t_{2},\cdots)r_{12}(1)=0, 
\en
which can be rewritten in terms of functionals $\kappa _a$ as follows:
\begin{align}
2\kappa_a \(AS_0\)
-i(1+J_{bc})\kappa_c \(AS_b\)
+i(1-J_{bc})\kappa _b \(AS_c\)=0 \quad \forall A\in 
\bA_{n-2}. 
\label{lasteq}
\end{align}
These equations can be viewed as a system of linear equations for
three constants (\ref{const}). 
Certainly, these equations are not explicit
since for every $A$ we have to 
perform the procedure of reducing to
PBW form. 
This huge 
system of homogeneous linear equations does not 
allow us to prove that the constants in question vanish;  
rather they reduce them to one constant.
Let us explain this point. 
First, it is clear that the equations (\ref{lasteq})
correspond to the following relation in 
$\bA_n$
$$
2 S_0S_a-i(1+J_{bc})S_bS_c+i(1-J_{bc})S_cS_b=0, 
$$
obtained by solving  \eqref{eq:quad1},
\eqref{eq:quad2} for $S_0S_a$.  
So, it is easy to see that all our 
equations including (\ref{lasteq})
are satisfied by the following
construction. Consider a linear 
functional $\kappa$ on 
$\bA_n$
such that $\kappa\(S_{\alpha_1}\cdots S_{\alpha_n}\)=0$ 
unless $\sum _{j=1}^n\bar{\alpha}_j=(0,0)$, and 
$$
\kappa (AS_0)=0. 
$$
Then all the requirements are satisfied by
\begin{align}
\kappa _{a}(A)=\kappa \(AS_a\). 
\nn
\end{align}
On the other hand the 
number of solutions to the system of linear equations
(\ref{lasteq}) for three 
constants cannot be bigger for arbitrary $\eta$
than it is for $\eta =0$. 
In the latter case the algebra is commutative
(see Appendix \ref{app:existence}, 
notably \eqref{eq:scale1}), 
and the equations (\ref{lasteq}) become
$$
\kappa _a^{\text{cl}}\(S_bA\)=\kappa _b^{\text{cl}}\(S_aA\)
$$
with additional condition
$\kappa _a^{\text{cl}}\(S_0A\)=0$. 
It is easy to see that this gives a  system of
three equations for three constants whose rank equals 2.

Thus we come to the conclusion:
\begin{lem}\label{lem:kappa}
Under the induction hypothesis, we have
\begin{align}
&\kappa _{\alpha_n,\cdots ,\alpha _1}=
\kappa \(S_{\alpha_n}\cdots S_{\alpha_1}\)\nn
\end{align}
where $\kappa$ is a linear functional on 
$\bA_n$
satisfying
\begin{align}
&\kappa\(S_{\alpha_n}\cdots S_{\alpha_1}\)=0
\ \ \text{ unless} \ \sum _{j=1}^n
\bar{\alpha}_j=(0,0)\nn\\
&\kappa (AS_0)=\kappa (S_0A)=0\nn
\end{align}
and as such is defined by one constant:
\begin{align}
&\kappa (S_3^{n})
\qquad \quad \quad \text{for $n$ even}, \nonumber \\
&\kappa(S_3^{n-2}S_1S_2)
\quad \ \text{for $n$ odd}. 
\nonumber 
\end{align}
\end{lem}
Unfortunately, we were not able to show that this remaining
constant equals zero.
The problem is still open. 

\section{Examples}

In this section we write down the Ansatz in 
the simplest cases $n=2,3$. 
We also consider the trigonometric limit. 

\subsection{The case $n=2$}

In the case $n=2$, $\Omega^{(1,2)}_2(t_1,t_2)$ can be readily 
found from \eqref{eq:trSa2}.
The function $h_{2}(t_{1}, t_{2})$ is given as follows:
\begin{eqnarray*}
h_{2}(t_{1}, t_{2})=
\frac{1}{4}
-
\frac{1}{4[t_{12}]}
\sum_{a=1}^{3}H_{a+1}(2t_{12})\, 
\sigma^{a}\otimes\sigma^{a}, 
\end{eqnarray*}
where 
\begin{eqnarray*}
H_{a+1}(2t):=
\frac{\varepsilon_{a}
\theta_{a+1}^2\theta_{a+1}(2\eta)\theta_{a+1}(2t)}
{4(\theta_{1}')^{2}\theta_{1}(2\eta)^2}
\left( 
\frac{\theta_{a+1}'(2t)}{\theta_{a+1}(2t)}\omega_{1}(t)+
\frac{\theta_{a+1}'(2\eta)}{\theta_{a+1}(2\eta)}\omega_{2}(t)
-4\pi i \,
\omega_{3}(t) \right). 
\end{eqnarray*}
This gives the formula for the nearest neighbor correlators 
of the inhomogeneous chain:
\bea
\langle\sigma^a_1\sigma^a_2\rangle=
-\frac{\theta_{1}(2\eta)}{\theta_{1}(2t)}H_{a+1}(2t),
\label{eq:n=2}
\ena
where $a=1,2,3$ and $t=t_{12}$. 
Noting that 
$H_{a+1}(2t)$ is odd in $t$,
we obtain 
$\langle\sigma^a_1\sigma^a_2\rangle=
 -\theta_{1}(2\eta)H_{a+1}'(0)/\theta_{1}'$ 
in the homogeneous limit $t\to0$, 
or more explicitly we have  
\bea
&&\langle\sigma^a_1\sigma^a_2\rangle
=-\frac{\ve_a\theta_{a+1}^2}
{8{\theta_1'}^3\theta_1(2\eta)}
\label{eq:n=2hom}
\\
&&\quad\times
\left(2\theta_{a+1}''(0)\theta_{a+1}(2\eta)
+\theta_{a+1}\theta_{a+1}'(2\eta)\frac{\partial}{\partial\eta}
-4\pi i\,
\theta_{a+1}\theta_{a+1}(2\eta)\frac{\partial}{\partial\tau}
\right)\omega_1(0). 
\nn
\ena

Let us check the formula 
\eqref{eq:n=2hom} against known results.  
As is well known, the XYZ Hamiltonian is obtained by differentiating 
the transfer matrix of the eight-vertex model
\be
T_L(t)=\tr\Bigl(R_{0L}(t)\cdots R_{01}(t)\Bigr)
\en
as
\bea
T_L(0)^{-1}T_L'(0)
=\sum_{j=1}^L\left(\sum_{a=1}^3 v_a'(0)\sigma^a_j\sigma^a_{j+1}\right)+Lv_0'(0),
\label{eq:R-Ham}
\ena
where $L$ is the length of the chain, and we have set
$\check{R}(t)=\sum_{\alpha=0}^3v_\alpha(t)\sigma^\alpha\otimes\sigma^\alpha$.
As it was mentioned already, the $R$ matrix \eqref{eq:R} 
is so normailized that in the thermodynamic limit $L\to\infty$ 
the free energy per site of the eight-vertex model is $0$. 
Therefore, taking the ground state average of \eqref{eq:R-Ham}, we obtain
\bea
&&\sum_{a=1}^3 I^a\langle \sigma^a_1\sigma^a_2\rangle
=-I^0, 
\label{eq:n=2eq1}
\ena
with $I^\alpha=v_\alpha'(0)\theta_1(2\eta)/\theta_1'$. Explicitly we have 
\bea
&&
I^a=\frac{\theta_{a+1}(2\eta)}{\theta_{a+1}}
\quad (a=1,2,3),
\label{eq:Ia}\\
&&
I^0=\frac{\theta_1(2\eta)}{\theta_1'}\frac{1}{4}
\omega_1(0).
\nn
\ena
The average over the normalized ground state has the property
$\delta\langle H_{XYZ} \rangle =\langle \delta H_{XYZ} \rangle$, 
where $\delta$ stands for the variation of the coefficients $I^a$.  
Hence we have in addition
\bea
&&\sum_{a=1}^3
\frac{\partial I^a}{\partial\eta}
\langle \sigma^a_1\sigma^a_2\rangle
=-
\frac{\partial I^0}{\partial\eta}, 
\label{eq:n=2eq2}
\\
&&\sum_{a=1}^3
\frac{\partial I^a}{\partial\tau}
\langle \sigma^a_1\sigma^a_2\rangle
=-
\frac{\partial I^0}{\partial\tau}.
\label{eq:n=2eq3}
\ena
The nearest neighbor correlators 
$\langle\sigma_1^a\sigma_2^a\rangle$ 
are completely determined by the linear equations 
\eqref{eq:n=2eq1}, \eqref{eq:n=2eq2}, \eqref{eq:n=2eq3}.
Using Riemann's identity and the heat equation
$4\pi i\partial\theta_\alpha(t|\tau)/\partial\tau
=\partial^2\theta_\alpha(t|\tau)/\partial t^2$, 
one can verify that our formula \eqref{eq:n=2hom} 
indeed gives the unique solution. 

\subsection{The case $n=3$}

Let us proceed to the next case $n=3$.  
Written in full, 
$h_{3}(t_{1}, t_{2}, t_{3})$ reads
\begin{eqnarray*}
h_{3}(t_{1}, t_{2}, t_{3})=
\frac{1}{8}-\frac{1}{16}
\frac{1}{[t_{12}][t_{13}][t_{23}]}
\sum_{(\alpha, \beta, \gamma)\not=(0, 0, 0) \atop 
{\bar{\alpha}+\bar{\beta}+\bar{\gamma}=\bar{0}}}
\hspace{-1em}
\sigma^{\alpha}\otimes\sigma^{\beta}\otimes\sigma^{\gamma}
\sum_{1\le j<k \le 3}
I_{\alpha, \beta, \gamma}^{(j,k)}(t_{1}, t_{2}, t_{3})\, . 
\end{eqnarray*}
The coefficients $I_{\alpha,\beta,\gamma}^{(j,k)}$ are given as follows: 
\begin{eqnarray*}
&& 
I_{0,1,1}^{(1,2)}=0, \\ 
&& 
I_{1,0,1}^{(1,2)}=
\frac{\theta_{2}}{\theta_{2}(2\eta)}\left\{
\frac{\theta_{4}(2t_{13})\theta_{3}(2t_{23})}
{\theta_{4}(2\eta)\theta_{3}}
H_{3}(2t_{12})+
\frac{\theta_{3}(2t_{13})\theta_{4}(2t_{23})}
{\theta_{3}(2\eta)\theta_{4}}
H_{4}(2t_{12})
\right\}, \\ 
&& 
I_{1,1,0}^{(1,2)}=
2[t_{13}][t_{23}]\, H_{2}(2t_{12}), \\ 
&& 
I_{1,2,3}^{(1,2)}=
(-i)\left\{ 
[t_{13}]\frac{\theta_{4}(2t_{23})}{\theta_{4}(2\eta)}
H_{2}(2t_{12})-
\frac{\theta_{2}\theta_{3}(2\eta)}{\theta_{2}(2\eta)\theta_{3}}
\frac{\theta_{4}(2t_{13})}{\theta_{4}(2\eta)}
[t_{23}]H_{3}(2t_{12})
\right\}, \\ 
&& 
I_{1,3,2}^{(1,2)}=
(-i)\left\{ 
\frac{\theta_{2}\theta_{4}(2\eta)}{\theta_{2}(2\eta)\theta_{4}}
\frac{\theta_{3}(2t_{13})}{\theta_{3}(2\eta)}[t_{23}]
H_{4}(2t_{12})-
[t_{13}]\frac{\theta_{3}(2t_{23})}{\theta_{3}(2\eta)}
H_{2}(2t_{12})
\right\}, 
\end{eqnarray*}
\begin{eqnarray*}
&& 
I_{0,1,1}^{(1,3)}=I_{1,1,0}^{(1,3)}=0, \\ 
&& 
I_{1,0,1}^{(1,3)}=
\frac{\theta_{3}^{2}(2\eta)\theta_{4}^{2}+\theta_{4}^{2}(2\eta)\theta_{3}^{2}}
{\theta_{3}\theta_{4}\theta_{3}(2\eta)\theta_{4}(2\eta)}
[t_{12}][t_{23}]
H_{2}(2t_{13}) \\ 
&& \hspace{4em} {}-
\frac{\theta_{2}}{\theta_{2}(2\eta)}
\left\{
\frac{\theta_{4}(2t_{12})\theta_{4}(2t_{23})}
{\theta_{4}(2\eta)\theta_{4}}
H_{3}(2t_{13})+
\frac{\theta_{3}(2t_{12})\theta_{3}(2t_{23})}
{\theta_{3}(2\eta)\theta_{3}}
H_{4}(2t_{13})
\right\}, \\ 
&& 
I_{1,2,3}^{(1,3)}=
(-i)\left\{
\frac{\theta_{2}}{\theta_{3}}
\frac{\theta_{3}(2t_{12})}{\theta_{2}(2\eta)}
[t_{23}]H_{4}(2t_{13})-
[t_{12}]
\frac{\theta_{4}}{\theta_{3}}
\frac{\theta_{3}(2t_{23})}{\theta_{4}(2\eta)}
H_{2}(2t_{13})
\right\}, \\ 
&& 
I_{1,3,2}^{(1,3)}=
(-i)\left\{
[t_{12}] 
\frac{\theta_{3}}{\theta_{4}}
\frac{\theta_{4}(2t_{23})}{\theta_{3}(2\eta)}
H_{2}(2t_{13})-
\frac{\theta_{2}}{\theta_{4}}
\frac{\theta_{4}(2t_{12})}{\theta_{2}(2\eta)}
[t_{23}]H_{3}(2t_{13})
\right\}, 
\end{eqnarray*}
\begin{eqnarray*}
&& 
I_{1,1,0}^{(2,3)}=0, \\ 
&& 
I_{1,0,1}^{(2,3)}=
\frac{\theta_{2}}{\theta_{2}(2\eta)}\left\{
\frac{\theta_{3}(2t_{12})\theta_{4}(2t_{13})}{\theta_{3}\theta_{4}(2\eta)}
H_{3}(2t_{23})+
\frac{\theta_{4}(2t_{12})\theta_{3}(2t_{13})}{\theta_{3}(2\eta)\theta_{4}}
H_{4}(2t_{23})
\right\}, \\ 
&& 
I_{0,1,1}^{(2,3)}=
2[t_{12}][t_{13}]\, H_{2}(2t_{23}), \\ 
&& 
I_{1,2,3}^{(2,3)}=
(-i)\left\{ 
\frac{\theta_{3}(2\eta)\theta_{4}}{\theta_{3}\theta_{4}(2\eta)}
\frac{\theta_{2}(2t_{13})}{\theta_{2}(2\eta)}
[t_{12}]H_{3}(2t_{23})-
[t_{13}]\frac{\theta_{2}(2t_{12})}{\theta_{2}(2\eta)}
H_{4}(2t_{23})
\right\}, \\ 
&& 
I_{1,3,2}^{(2,3)}=
(-i)\left\{ 
[t_{13}]\frac{\theta_{2}(2t_{12})}{\theta_{2}(2\eta)}
H_{3}(2t_{2,3})-
\frac{\theta_{3}\theta_{4}(2\eta)}{\theta_{3}(2\eta)\theta_{4}}
\frac{\theta_{2}(2t_{13})}{\theta_{2}(2\eta)}[t_{12}]
H_{4}(2t_{23})
\right\}.
\end{eqnarray*}
The rest are given by the cyclic change $1\to2\to3\to1$ 
of the indices $\alpha, \beta, \gamma$ in 
$I_{\alpha, \beta, \gamma}^{(j,k)}$
with the change $2\to3\to4\to2$ of 
the indices in $\theta_{a}$ and $H_{a}$. 
The correlators of the inhomogeneous chain are 
\begin{eqnarray*}
\langle \sigma_{1}^{\alpha}\sigma_{2}^{\beta}\sigma_{3}^{\gamma} \rangle 
=-\frac{1}{2[t_{12}][t_{13}][t_{23}]}
\sum_{1 \le j<k \le 3}I_{\alpha,\beta,\gamma}^{(j, k)}(t_{1}, t_{2}, t_{3}). 
\end{eqnarray*}


With the abbreviation 
$H_{a}'=H_{a}'(0)$ and $H_{a}'''=H_{a}'''(0)$, 
we obtain a new formula for 
the next nearest neighbor correlators 
for the homogeneous chain 
\begin{eqnarray*}
&& 
\langle \sigma_{1}^{a} \sigma_{3}^{a} \rangle=
{}-\frac{1}{4}\frac{\theta_{1}(2\eta)}{\theta_{1}'}\Bigl\{ 
2\frac{\theta_{b+1}^{2}(2\eta)\theta_{c+1}^{2}+
       \theta_{c+1}^{2}(2\eta)\theta_{b+1}^{2}}
{\theta_{b+1}\theta_{c+1}\theta_{b+1}(2\eta)\theta_{c+1}(2\eta)}\, 
H_{a+1}' \\ 
&& \hspace{6em} {}+
\left(\frac{\theta_{1}(2\eta)}{\theta_{1}'}\right)^{2}
\frac{\theta_{a+1}}{\theta_{a+1}(2\eta)}
\Bigl\{
\frac{\theta_{c+1}}{\theta_{c+1}(2\eta)}
(\frac{\theta_{b+1}''}{\theta_{b+1}}H_{b+1}'+
\frac{2\theta_{c+1}''}{\theta_{c+1}}H_{b+1}'-H_{b+1}''') \\ 
&& \hspace{15em} {}+
\frac{\theta_{b+1}}{\theta_{b+1}(2\eta)}
(\frac{\theta_{c+1}''}{\theta_{c+1}}H_{c+1}'+
\frac{2\theta_{b+1}''}{\theta_{b+1}}H_{c+1}'-H_{c+1}''')
\Bigr\}
\Bigr\}. 
\end{eqnarray*}
We have in addition 
\begin{eqnarray*}
\langle \sigma_{1}^{a}\sigma_{2}^{b}\sigma_{3}^{c} \rangle=0,  
\qquad 
\langle \sigma_{1}^{c}\sigma_{2}^{b}\sigma_{3}^{a} \rangle=0 . 
\end{eqnarray*}
In both formulas, $(a,b,c)=(1,2,3), (2,3,1), (3,1,2)$. 

M. Lashkevich communicated to us a program for 
numerically calculating correlation functions 
from the integral formula of \cite{LP1,LP2}. 
For $n=2$ and $n=3$, 
we found 
agreement between their results and ours
to within the precision $10^{-4}$.

\subsection{Trigonometric limit}

Finally we briefly touch upon the trigonometric limit,
and discuss
how various quantities which appear 
in \eqref{eq:n=2} are related to the trigonometric counterpart. 

First we consider the limit to the massive regime. 
For this purpose, it is convenient to rewrite the $R$ matrix 
in terms of the parameters $t',\eta',\tau'$ in \eqref{eq:modular}
as
\be
R(t)=\frac{\rho'(t')}{[t'+\eta']'}(U\otimes U)  r'(t') (U\otimes U)^{-1},
\en
where $[t']'=\theta_1(2t'|\tau')/\theta_1(2\eta'|\tau')$, 
$r'(t')$ is obtained from \eqref{eq:r} by replacing $t,\eta,\tau$ by
$t',\eta',\tau'$, and $U=
\begin{pmatrix}1&1\\1&-1\\ \end{pmatrix}$. 
In the limit $\tau' \to +i\infty$ while 
keeping $\lambda=t'/\eta'$ and $\nu=2\eta'$ fixed, 
the $R$ matrix tends to  
\bea
&& 
R_{\rm XXZ}(\lambda)=\rho_{\rm XXZ}(\lambda)
\frac{r_{\rm XXZ}(\lambda)}{[\lambda+1]_{XXZ}}.
\label{eq:R-matrix-XXZ}
\ena
In the above, $[\la]_{XXZ}=\sin\pi\nu\la/\sin\pi\nu$, and
\bea
&& 
r_{\rm XXZ}(\lambda)=
\frac{1}{2}
\Bigl(
\frac{\sin{(\lambda+1/2)\pi\nu}}{\sin{\pi \nu/2}}
\sigma^{0}\otimes \sigma^{0}
\label{eq:trig-R} \\ 
&&\quad
+
\sigma^{1}\otimes\sigma^{1}+\sigma^{2}\otimes\sigma^{2}+
\frac{\cos{(\lambda+1/2)\pi\nu}}{\cos{\pi \nu/2}}
\sigma^{3}\otimes\sigma^{3}
\Bigr), 
\nn\\
&&
\rho_{\rm XXZ}(\lambda)=
{}-\zeta\frac{(q^{2}\zeta^{2})_{\infty}(\zeta^{-2})_{\infty}}
{(q^{2}\zeta^{-2})_{\infty}(\zeta^{2})_{\infty}}, 
\label{eq:def-rho-massive}
\end{eqnarray}
where $\zeta=e^{\pi i\nu \lambda}$, $q=e^{ \pi i \nu}$,
$(x)_\infty=\prod_{j=0}^\infty(1-q^{4j}x)$. 
It is easy to see that 
\begin{eqnarray*}
&& 
\omega_{1}(t) \to 
\frac{8\pi}{\sin{\pi \nu}}
\omega(\lambda), \quad
\omega_{2}(t)\to 
\frac{8\pi}{\sin{\pi \nu}}
\tilde{\omega}(\lambda), \quad 
\omega_{3}(t)\to 0,
\end{eqnarray*}
where 
the functions $\omega(\lambda), \tilde{\omega}(\lambda)$ are 
given by \cite{BJMST2}, eqs.(13.2)--(13.5) for the massive regime. 
Hence the limit of \eqref{eq:n=2} becomes
\begin{eqnarray}
&& \hspace{1em}
\lim_{\tau'\to i\infty}{\langle \sigma^{3}_{1}\sigma^{3}_{2} \rangle}=
{}-4\left( 
\frac{q+q^{-1}}{(q-q^{-1})^{2}}\omega(\lambda)+
\frac{\zeta+\zeta^{-1}}{(q-q^{-1})(\zeta-\zeta^{-1})}
\tilde{\omega}(\lambda)
\right), 
\label{eq:trig-limit} \\ 
&& 
\lim_{\tau'\to i\infty}{\langle \sigma^{1}_{1}\sigma^{1}_{2} \rangle}=
\lim_{\tau'\to i\infty}{\langle \sigma^{2}_{1}\sigma^{2}_{2} \rangle}=
2\left( 
\frac{\zeta+\zeta^{-1}}{(q-q^{-1})^{2}}\omega(\lambda)+
\frac{q+q^{-1}}{(q-q^{-1})(\zeta-\zeta^{-1})} 
\tilde{\omega}(\lambda) \right),
\nonumber
\end{eqnarray}
which 
reproduces the formulas 
in the massive regime
(see \cite{BJMST2}, Example in Section 3).

Second let us consider the limit to the massless regime. 
We set 
\begin{eqnarray*}
\tau=-\frac{1}{\pi i}r, \quad 
\eta=-\frac{\nu}{2\pi i}r, \quad 
t=-\frac{\nu \lambda}{2\pi i}r 
\end{eqnarray*}
for a constant $\nu \,\, (0<\nu<1)$ and 
take the limit $r \downarrow 0$ with $\nu$ and 
$\lambda$ 
fixed. 

The limit of the $R$ matrix is given by the same formula 
\eqref{eq:R-matrix-XXZ}--\eqref{eq:trig-R}, 
with $\rho_{\rm XXZ}(\la)$ being replaced by
\be
\rho_{\rm XXZ}(\lambda)=-\frac{S_{2}(-\lambda)S_{2}(1+\lambda)}
{S_{2}(\lambda)S_{2}(1-\lambda)}.
\en
Here $S_{2}(x)=S_{2}(x \, | 2, 1/\nu)$ 
signifies the double sine function. 
In the limit we have 
\begin{eqnarray*}
r\omega_{1}(t) \to 
{}-\frac{8\pi^{2}i}{\sin{\pi \nu}}\omega(\lambda), \qquad 
r\omega_{2}(t)\to 
{}-\frac{8\pi^{2}i}{\sin{\pi \nu}}\tilde{\omega}(\lambda),
\end{eqnarray*}
where now $\omega(\lambda)$ and $\tilde{\omega}(\lambda)$ 
stand for the functions given by 
 \cite{BJMST2}, eqs.(13.2)--(13.5) for the massless regime. 
Moreover we have 
\begin{eqnarray*}
re^{\frac{\pi^{2}}{r}}
\left( 
t\frac{\partial}{\partial t}+
\eta\frac{\partial}{\partial \eta}+
\tau\frac{\partial}{\partial \tau} \right)\log{\varphi} 
\to 0. 
\end{eqnarray*}
{}From the formulas above, we see that 
in the massless limit 
the function $h_{2}(t_{1}, t_{2})$ tends to 
the solution $h_{2}(\lambda_{1}, \lambda_{2})$ 
of the reduced qKZ equation given in \cite{BJMST2}. 

\appendix

\section{Existence of $\Tr_\la$}\label{app:existence}

For every finite-dimensional representation of the 
Sklyanin algebra $\A$, we can define the trace, 
which is a functional on $\A$ whose main property
is cyclicity. 
In order to formulate our Anstaz for correlation functions,
we need an analytic continuation of this functional 
with respect to the dimension. 
We denote this analytic continuation by ${\rm Tr}_\lambda A$,
where $A\in\A$ and $\lambda=k+1$ for $\pi^{(k)}(A)$.
In Section \ref{subsec:trace}, 
we presented the formulas for 
${\rm Tr}_\lambda S_\alpha$,  
${\rm Tr}_\lambda S_\alpha^2$ 
($\alpha=0,1,2,3$).
In this appendix, we discuss the general case. In fact, we prove
that for generic parameters $J_1,J_2,J_3$, the definition of
${\rm Tr}_\lambda A$ for general $A\in\A$ can be reduced to these known
cases.

Consider the polynomial ring $\Fb=\Kb[K_0,K_2]$ with $\Kb=\C(J_1,J_2,J_3)$.
Here, we consider $K_0,K_2,J_1,J_2,J_3$ as variables,
whereas they are parameterized by $\tau$, $\eta$ and $\lambda$ 
in Section \ref{sec:2} and Appendix \ref{app:cycles}.  
We use the parameterization in order to define finite dimensional
representations. The discussion in this appendix is mainly concerned with
the algebraic relations in the Sklyanin algebra only.

We denote by $\Ab$ the Sklyanin algebra defined over the field $\Kb$.
It is a graded vector space, 
\be
\Ab=\oplus_{n=0}^\infty\Ab_n,
\quad \dim \Ab_n<\infty. 
\en

Multiplication by the central elements (\ref{eq:Cas})
endows $\Ab$ with an $\Fb$-algebra structure.  
Suppose we try to define some $\Fb$-linear functional $\Tr$ 
on $\Ab$ which satisfies 
cyclicity ${\rm Tr}\,(AB)={\rm Tr}\,(BA)$. 
Then the question is, 
for how many independent elements of $\Ab$ 
this functional should be defined.
In other words, describe the $\Fb$-module
\be
\Hb=\Ab/\Ab'
\en
where
\be
\Ab'=\sum_{\alpha=0}^3[S_\alpha,\Ab].
\en
Note that $\Hb=\oplus_{n=0}^\infty\Hb_n$ where
$\Hb_n=\Ab_n/\Ab_n'$, 
$\Ab_n'=\sum_{\alpha=0}^3[S_\alpha,\Ab_{n-1}]$.

We prove
\begin{thm}\label{SEVEN}
The $\Fb$-module $\Hb$ is a rank $7$ free module generated by the
monomials
\bea
(m_i)_{1\leq i\leq 7}=(1,S_0,S_1,S_2,S_3,S_0^2,S_3^2).
\ena
\end{thm}

The $\Fb$-linear independence of these elements follows from
\eqref{eq:trSa1}--\eqref{eq:fa2}. 
Indeed, suppose there is a relation $\sum_{i=1}^7 c_im_i=0$
with $c_i\in\Fb$. 
The sum over elements of even degree and of 
odd degree must vanish separately. 
Specialize $J_i$ to the value \eqref{eq:J2}
with $\eta\not\in\Q+\Q\tau$, $\Im\eta>0$, 
and take the trace of both sides on the representation 
$\V^{(k)}$ for $k\in\Z_{\ge 0}$.   
By Lemma \ref{lem:unique}, it follows that 
$c_i=0$ except for $i=3,4,5$. 
To see that the latter vanish, it is enough to 
apply the automorphisms $\iota^1,\iota^3$ 
(see \eqref{eq:Liota1},\eqref{eq:Liota3} and two lines above) 
and take the trace.  

Let us prove the spanning property. 

Consider the tensor algebra $\Tb$ over the field $\Kb$
generated by four independent variables
$S_0$, $S_1$, $S_2$ and $S_3$. Set $\Rb=\Tb[K_0,K_2]$. It is a graded
algebra: $\Rb=\oplus_{n=0}^\infty \Rb_n$, where we have
${\rm dim}_{\Kb}\Rb_n<\infty$. We have the isomorphism of $\Kb$-vector spaces
\bea\label{VECISO}
\Hb_n\simeq\Rb_n/\Biggl(\sum\Rb_{n-2}\bigl
([S_0,S_a]-iJ_{b,c}(S_bS_c+S_cS_b)\bigr)\\
+\sum\Rb_{n-2}\bigl([S_b,S_c]-i(S_0S_a+S_aS_0)\bigr)
+\sum[S_a,\Rb_{n-1}]\nonumber\\
+\Rb_{n-2}\Bigl(\sum_{\alpha=0}^3S_\alpha^2-K_0\Bigr)
+\Rb_{n-2}\Bigl(\sum_{a=1}^3J_aS_a^2-K_2\Bigr)
\Biggr).\nonumber
\ena
The $\Kb$-vector space $\Rb_n$ is spanned by the monomials of the form
$K_0^{m_0}K_2^{m_2}S_{\alpha_1}\cdots S_{\alpha_l}$ where
$2m_0+2m_2+l=n$. The relations which define $\Hb_n$ in (\ref{VECISO})
are linear relations for these monomials. For each $n$ the coefficients of
these linear relations form a matrix $\mathcal{M}_n$ with entries in $\Kb$.

The spanning property is clear for $n=0,1$. Suppose that $n\geq2$.
Divide the set of monomials of degree $n$ into two groups:
the first group is the monomials such that the part 
$S_{\alpha_1}\cdots S_{\alpha_l}$
is equal to one of $m_i$ $(1\leq i\leq 7)$ and the second group is the rest.
The matrix $\mathcal{M}_n$ is divided into two blocks 
$\mathcal{M}_n=(\mathcal{M}'_n,\mathcal{M}''_n)$
where $\mathcal{M}'_n$ (resp., $\mathcal{M}''_n$) 
corresponds to the first (resp., the second)
group of monomials. It suffices to show that 
the rank of $\mathcal{M}''_n$ is equal to
the cardinality of the second group.

The proof of this statement exploits the classical limit:
\bea
J_a=1-\varepsilon^2j_a,\quad\varepsilon\rightarrow0.
\ena
We introduce new variables $s_\alpha$ $(\alpha=0,1,2,3)$ and $k_0,k_1$ by
\bea
&&
S_0=\varepsilon s_0,\quad S_a=s_a\quad(a=1,2,3),
\label{eq:scale1}
\\
&&K_0=k_0,\quad K_2=k_0-\varepsilon^2k_1.
\label{eq:scale2}
\ena
In Appendix \ref{app:cycles}, the classical limit is taken as
$\eta\rightarrow0$ instead of $\varepsilon\rightarrow0$.
In this appendix, we avoid the parametrization by $\tau$, $\eta$ and $\lambda$
in order to simplify the argument.

In the limit $\varepsilon\rightarrow0$, we have
\bea\label{GRASS}
[s_\alpha,s_\beta]=i\varepsilon\{s_\alpha,s_\beta\}+O(\varepsilon^2),
\ena
where the Poisson bracket is defined by
\bea
\{s_0,s_a\}&=&2j_{b,c}s_bs_c,
\label{eq:Poisson1}
\\
\{s_b,s_c\}&=&2s_0s_a.
\label{eq:Poisson2}
\ena
Here $j_{a,b}=-j_{b,a}=j_a-j_b$ and $(a,b,c)$ runs over
cyclic permutations of $(1,2,3)$. In the classical limit, the variables $s_a$
become commutative. 
Let 
$\Kc=\C(j_1,j_2,j_3)$
denote the field of rational functions in $j_a$, and 
let $\Fc=\Kc[k_0,k_1]$, 
$\Ac=\Kc[s_0,s_1,s_2,s_3]$ denote 
the polynomial ring in indeterminates 
$k_0,k_1$ and $s_\alpha$ ($0\le \alpha\le 3$), respectively. 
The Casimir relations (\ref{eq:Cas}) become
the algebraic relations
\bea
&&s_1^2+s_2^2+s_3^2=k_0,\label{QUAD1}\\
&&s_0^2+j_1s_1^2+j_2s_2^2+j_3s_3^2=k_1\label{QUAD2}
\ena
in $\Kc[s_0,s_1,s_2,s_3,k_0,k_1]$:
\be
\Ab^{\rm cl}&\simeq&
\Kc[s_0,s_1,s_2,s_3,k_0,k_1]/
(s_1^2+s_2^2+s_3^2-k_0,s_0^2+j_1s_1^2+j_2s_2^2+j_3s_3^2-k_1).
\en
This makes $\Ac$ an $\Fc$-algebra. 
The algebra $\Ac$ is graded as well: 
$\Ac=\oplus_{n=0}^\infty {\bf A}_{n}^{\rm cl}$.
Let $(\Ac)'_n\subset {\bf A}_{n}^{\rm cl}$ 
be the limit (in the appropriate Grassmannian such that we consider
$s_\alpha$ as commutative variables) of 
$\Ab'_n$ as $\varepsilon\to 0$. 
Then, we see 
$\sum_{\alpha=0}^3\{s_\alpha, {\bf A}_{n-1}^{\rm cl}\}\subset (\Ac)'_n$
{}from (\ref{GRASS}). In order to show the spanning property of 
$(m_i)_{1\leq i\leq 7}$ in $\Hb$, 
it is therefore sufficient to show the spanning property for 
\bea\label{CLAM}
(m_i^{\rm cl})_{1\leq i\leq 7}=(1,s_0,s_1,s_2,s_3,s_0^2,s_3^2).
\ena
in $\Hc=\oplus_{n=0}^\infty {\bf H}_{n}^{\rm cl}$, where
\bea\label{QQ}
\Hb^{\rm cl}_n={\bf A}_n^{\rm cl}/
\sum_{\alpha=0}^3\{s_\alpha,{\bf A}_{n-1}^{\rm cl}{}\}. 
\ena

In conclusion, Theorem \ref{SEVEN} follows from
\begin{prop}\label{CLAS}
The $\Fc$-module $\Hc$ is a rank $7$ free module generated by
the monomials $(\ref{CLAM})$.
\end{prop}

\begin{proof}
The $\Fc$-linear independence of $m^{\rm cl}_i$ follows 
{}from the same argument as in the quantum case. 
In place of trace on $\V^{(k)}$ we use 
the non-degenerate pairing between cycles and cocycles 
given in Appendix \ref{app:cycles}. 

We prove the spanning property.
Define
\be
\nabla_0&=&j_{2,3}s_2s_3\frac{\partial}{\partial s_1}+
j_{3,1}s_3s_1\frac{\partial}{\partial s_2}+
j_{1,2}s_1s_2\frac{\partial}{\partial s_3},\\
\nabla_a&=&-j_{b,c}s_bs_c\frac{\partial}{\partial s_0}+
s_0s_c\frac{\partial}{\partial s_b}
-s_0s_b\frac{\partial}{\partial s_c},
\en
where $(a,b,c)=(1,2,3)$, $(2,3,1)$, $(3,1,2)$. We have
$\nabla_\alpha P=\frac12\{s_\alpha,P\}$. These are $\Fc$-linear.

We want to show that modulo $\sum_{\alpha=0}^3\nabla_\alpha\Ac$ any
monomial
$s_0^{m_0}s_1^{m_1}s_2^{m_2}s_3^{m_3}$ can be reduced to an element in
$\Fc\cdot1+\sum_{\alpha=0}^3\Fc\cdot s_\alpha+
\sum_{\alpha=0}^3\Fc\cdot s_\alpha^2$.
Set
\be
&&\Ac{}^{\scriptscriptstyle[-1]}=\sum_{0\leq \alpha<\beta\leq3}
s_\alpha s_\beta\Ac, 
\\
&&
\Hc{}^{\scriptscriptstyle[-1]}=\Ac{}^{\scriptscriptstyle[-1]}/
\sum_{\alpha=0}^3\nabla_\alpha\Ac.
\en
We also denote $\Hc{}^{\scriptscriptstyle[0]}=\Hc$
and $\Ac{}^{\scriptscriptstyle[0]}=\Ac$.

Since
\be
\Ac=\Bigl(\sum_{\alpha=0}^3\sum_{n=0}^\infty\Kc s_\alpha^n\Bigr)
\oplus\Ac{}^{\scriptscriptstyle[-1]},
\en
by using (\ref{QUAD1}) and (\ref{QUAD2})
the above statement follows from the following:
\bea
\Hc{}^{\scriptscriptstyle[-1]}&=&\Fc k_0s_0+\sum_{a=1}^3\Fc(k_1-j_ak_0)s_a
+\sum_{a=1}^3\Fc(k_1-j_ak_0)s_a^2.\label{DSTAR}
\ena
Note that
\be
k_1-j_ak_0=s_0^2+j_{b,a}s_b^2+j_{c,a}s_c^2,
\en
and therefore we have $k_0s_0$, $(k_1-j_ak_0)s_a$,
$(k_1-j_ak_0)s_a^2\in\Ac{}^{\scriptscriptstyle[-1]}$.

Let us prove (\ref{DSTAR}). Suppose that a monomial
$m=s_0^{n_0}s_1^{n_1}s_2^{n_2}s_3^{n_3}\in\Hc{}^{\scriptscriptstyle[-1]}$
is such that
$\sharp\{a|n_a\in2\Z+1\}\geq2$: e.g., $n_0,n_1\in2\Z+1$.
By using the Casimir relations (\ref{QUAD1}) and (\ref{QUAD2})
we can replace $s_0^2$ and $s_1^2$ with $s_2^2$ and $s_3^2$. Therefore, we have
\be
m\in s_0s_1\Fc[s_2,s_3].
\en
Since
\be
\nabla_2(s_2^js_3^k)&=&ks_0s_1s_2^js_3^{k-1},
\en
we have
\be
m=0.
\en

Next consider the case $\sharp\{a|n_a\in2\Z+1\}=1$.
Suppose $n_0\in2\Z+1$. Then, we have 
\be
m\in s_0s_2^2\Fc[s_2^2,s_3^2]+s_0s_3^2\Fc[s_2^2,s_3^2].
\en
Since
\be
\nabla_1(s_1^is_2^ls_3^k)=ls_0s_1^i
s_2^{l-1}s_3^{k+1}-ks_0s_1^is_2^{l+1}s_3^{k-1},
\en
the monomial $m$ belongs to $\Kc s_0s_2^{2j}\subset\Hc{}^{\scriptscriptstyle[-1]}$
where $2j+1=n_0+n_1+n_2+n_3$. Similarly, we see that
\be
k_0^js_0\in\Kc s_0s_2^{2j}.
\en
Since $k_0^js_0$ is a non-zero element in $\Hc{}^{\scriptscriptstyle[-1]}$, we have
\be
m\in \Kc k_0^js_0.
\en
The case $n_a\in2\Z+1$ $(a=1,2,3)$ is similar.

The remaining case is $n_0,n_1,n_2,n_3\in2\Z$ and
$\sharp\{a|n_a>0\}\geq2$. We have ${\rm deg}\,m\geq4$.
Note that
\be
\Hc{}^{\scriptscriptstyle[-1]}_4=\sum_{0\leq i<j\leq3}\Kc s_i^2s_j^2.
\en
We have the following relations in $\Hc{}^{\scriptscriptstyle[-1]}_4$:
\be
\nabla_1(s_0s_2s_3)&=&-j_{2,3}s_2^2s_3^2+s_0^2s_3^2-s_0^2s_2^2,\\
\nabla_2(s_0s_3s_1)&=&-j_{3,1}s_3^2s_1^2+s_0^2s_1^2-s_0^2s_3^2,\\
\nabla_3(s_0s_1s_2)&=&-j_{1,2}s_1^2s_2^2+s_0^2s_2^2-s_0^2s_1^2.
\en
Therefore, we have ${\rm dim}_\Kc\Hc{}^{\scriptscriptstyle[-1]}_4\leq3$.
On the other hand $(k_1-j_ak_0)s_a^2\in\Hc{}^{\scriptscriptstyle[-1]}_4$ for
$a=1,2,3$, and they are $\Kc$-linearly independent.
Therefore, we have
\be
\Hc{}^{\scriptscriptstyle[-1]}_4=\oplus_{a=1}^3\Kc(k_1-j_ak_0)s_a^2.
\en

Observe that for $(a,b,c)=(1,2,3)$, $(2,3,1)$, $(3,1,2)$ we have
\be
\nabla_a(s_0^{n_0+1}s_1^{n_1}s_2^{n_2}s_3^{n_3})
=-(n_0+1)j_{b,c}s_0^{n_0}s_a^{n_a}s_b^{n_b+1}s_c^{n_c+1}\\
+n_bs_0^{n_0+2}s_a^{n_a}s_b^{n_b-1}s_c^{n_c+1}
-n_cs_0^{n_0+2}s_a^{n_a}s_b^{n_b+1}s_c^{n_c-1}.
\en
We can increase the power in $s_0$ by rewriting
$s_0^{n_0}s_a^{n_a}s_b^{n_b+1}s_c^{n_c+1}$ in terms of
$s_0^{n_0+2}s_a^{n_a}s_b^{n_b-1}s_c^{n_c+1}$
and $s_0^{n_0+2}s_a^{n_a}s_b^{n_b+1}s_c^{n_c-1}$.
Thus, we see that the $\Kc$ vector space
$\Hc{}^{\scriptscriptstyle[-1]}_{2j}$ $(j\geq3)$ is spanned by
\be
s_0^{2d}s_a^{2(j-d)}\quad(a=1,2,3;1\leq d\leq j-1).
\en
On the other hand, the space $\Hc{}^{\scriptscriptstyle[-1]}_{2j}$ contains
$\Kc$-linearly independent elements
\bea\label{TSTAR}
k_0^{d}k_1^{j-d-2}(k_1-j_ak_0)s_a^2\quad
(a=1,2,3;0\leq d\leq j-2).
\ena
Therefore, the elements (\ref{TSTAR}) span $\Hc{}^{\scriptscriptstyle[-1]}_{2j}$.
\end{proof}

In the rest of this appendix, 
we explain a mathematical background of
Proposition \ref{CLAS}, which is the
de Rham cohomology of the affine algebraic variety defined
by the two quadrics (\ref{QUAD1}) and (\ref{QUAD2}). Although our proof is independent, the statement of Proposition \ref{CLAS} is closely related to a result of K. Saito.

We set
\be
M=\{(j_1,j_2,j_3)\in\C^3|j_a\not=j_b\hbox{ for }a\not=b\}.
\en
Let
\bea
\varphi:X=\C^4\times M\rightarrow Y=\C^2\times M
\ena
be the mapping such that $\varphi=(\varphi_1,\ldots,\varphi_5)$
and for $({\bf s},{\bf j})=(s_0,s_1,s_2,s_3,j_1,j_2,j_3)\in\C^4\times M$
we have
\bea
\varphi_1({\bf s},{\bf j})&=&s_1^2+s_2^2+s_3^2,\\
\varphi_2({\bf s},{\bf j})&=&s_0^2+j_1s_1^2+j_2s_2^2+j_3s_3^2,\\
\varphi_3({\bf s},{\bf j})&=&j_1,\\
\varphi_4({\bf s},{\bf j})&=&j_2,\\
\varphi_5({\bf s},{\bf j})&=&j_3.
\ena
The critical set $C\subset\C^4\times M$ of 
this mapping is given by the equation
\be
d\varphi_1\wedge\cdots\wedge d\varphi_5=0.
\en
We have
\be
C=\bigcup_{\alpha=0}^3\{({\bf s},{\bf j})|s_\beta=0\hbox{ for }\beta\not=\alpha\}.
\en
We have the commutative diagram,
\begin{eqnarray*}
\begin{CD}
X @. \quad @. \supset @. \quad @. C \\ 
@ VVV @. @. @. @ VVV \\ 
Y @. \quad @. \supset @. \quad @. D,
\end{CD} 
\end{eqnarray*}
where the discriminant set $D$ is given by
\bea
D&=&\{(k_0,k_1,j_1,j_2,j_3)\in\C^2\times M|\Delta(k_0,k_1,j_1,j_2,j_3)=0\},\\
\Delta&=&k_0\prod_{a=1}^3(k_1-j_ak_0).
\ena
The inverse image $\varphi^{-1}(0)$ is called the simple elliptic singularity
of type $\tilde D_5$ \cite{Sa}. If $y\in Y$ does not belong to $D$,
the inverse image $X_y=\varphi^{-1}(y)$ is a
non-singular affine complex surface and is 
called a smoothing of the singularity.
The mapping
\bea
\varphi|_{X-\varphi^{-1}(D)}:
X-\varphi^{-1}(D)\rightarrow Y- D
\ena
is a locally topologically trivial fiber space, and the homology group
is of rank 7: $H_2(X_y,\Z)=\Z^7$. In Appendix \ref{app:cycles},
we construct cycles in $H_2(X_y,\Z)$.
(In \cite{Sa3}, K. Saito defined the extended affine root systems.
The homology group $H_2(X_y,\Z)$ is isomorphic to $D^{(1,1)}_5$
in his classification. We have not identified our cycles in
$D^{(1,1)}_5$.)

Let $\Omega^p_X$ be the sheaf of $\Oc_X$ modules
consisting of germs of holomorphic $p$ forms on $X$, and $\Omega^p_{X/Y}$
the quotient sheaf
\bea
\Omega^p_{X/Y}=
\Omega^p_X/\sum_{i=1}^5d\varphi_i\wedge\Omega^{p-1}_X.
\ena
The relative de Rham complex $(\Omega^{\bullet}_{X/Y},d_{X/Y})$
is defined by the commutative diagram
\be
d_{X/Y}&:&\Omega^p_{X/Y}\rightarrow \Omega^{p+1}_{X/Y}\\
&&\hskip10pt\uparrow\hskip35pt\uparrow\\
d&:&\hskip5pt\Omega^p_X\hskip5pt\rightarrow\hskip3pt\Omega^{p+1}_X.
\en
This is an exact sequence of $\Oc_X$ modules.
The following is K. Saito's result \cite{Sa4}.
\begin{thm}\label{SAITO}
The cohomology group
\be
H^2(\varphi_*(\Omega^{\bullet}_{X/Y}))=
{\rm Ker}(\varphi_*(\Omega^2_{X/Y})
\buildrel{d_{X/Y}}\over\rightarrow\varphi_*(\Omega^3_{X/Y}))
/{\rm Im}\,(\varphi_*(\Omega^1_{X/Y})
\buildrel{d_{X/Y}}\over\rightarrow\varphi_*(\Omega^2_{X/Y}))
\en
is an $\Oc_Y$ locally free module of rank $7$.
\end{thm}
We connect the above algebro-geometric setting to ours.
The following proposition is a corollary to Theorem \ref{SAITO}.
Here we give a proof in the line of this appendix without
using Saito's result.
\begin{prop}\label{PROP}
Consider a complex of $\Fc$-modules:
\be
Z^p&=&\bigoplus_{0\leq \alpha_1<\cdots<\alpha_p\leq 3}\Ac ds_{\alpha_1}\wedge\cdots\wedge ds_{\alpha_p},\\
\bar Z^p&=&Z^p/\sum_{j=1,2}d\varphi_j\wedge Z^{p-1}.
\en
The cohomology group $\Hc{}^{\scriptscriptstyle[-2]}\buildrel{\rm def}\over={\rm Ker}(\bar Z^2
\buildrel d\over\rightarrow\bar Z^3)
/{\rm Im}(\bar Z^1\buildrel d\over\rightarrow\bar Z^2)$ is a $\Fc$-free module
of rank $7$, where the action of $k_0$ $($resp., $k_1$$)$
is given by the multiplication of $\varphi_1$ $($resp., $\varphi_2$$)$.
\end{prop}
\begin{proof} The key idea of the proof is to identify
$\Ac{}^{\scriptscriptstyle[-1]}$ with $\bar Z^2$.

Let us introduce a holomorphic section $\omega$ of the sheaf $\Omega^2_{X/Y}\Big|_{X-C}$.
\be
&&\omega=\frac{ds_1\wedge ds_2}{s_0s_3}=\frac{ds_2\wedge ds_3}{s_0s_1}
=\frac{ds_3\wedge ds_1}{s_0s_2}\\
&&\quad=\frac{ds_0\wedge ds_1}{j_{2,3}s_2s_3}
=\frac{ds_0\wedge ds_2}{j_{3,1}s_3s_1}
=\frac{ds_0\wedge ds_3}{j_{1,2}s_1s_2}
\en
Consider the $\Fc$-module
\be
\tAc
\buildrel{\rm def}\over=
\Kc\otimes_{\C[j_1,j_2,j_3]}\C[s_0,s_1,s_2,s_3,j_1,j_2,j_3]\omega
\en
We have a canonical isomorphism of $\Fc$-modules
\be
\Ac\simeq\tAc,
\en
sending $P\in\Ac$ to $P\omega\in\tAc$.
It is easy to see that
\be
\Ac{}^{\scriptscriptstyle[-1]}\omega=\bar Z^2.
\en
We have
\bea\label{DIFPOI}
d(Pds_\alpha)=-\nabla_\alpha(P)\omega
\ena
for $P\in\Ac$. From (\ref{DIFPOI}) we have
\be
{\rm Im}(\bar Z^1\buildrel d\over\rightarrow\bar Z^2)
=\left(\sum_{\alpha=0}^3\nabla_\alpha\Ac\right)\omega.
\en
We have already constructed the $\Fc$-bases of the modules
$\Ac/\sum_{\alpha=0}^3\nabla_\alpha\Ac$ and 
$\Ac{}^{\scriptscriptstyle[-1]}/\sum_{\alpha=0}^3\nabla_\alpha\Ac$.
We will construct a basis of
${\rm Ker}(\bar Z^2\buildrel d\over\rightarrow\bar Z^3)/
{\rm Im}(\bar Z^1\buildrel d\over\rightarrow\bar Z^2)$.
 First, observe that
\be
\bar Z^3=\bigoplus_{n=0}^\infty\Kc s_0^nds_1\wedge ds_2\wedge ds_3
+\sum_{a=1}^3\bigoplus_{n=0}^\infty\Kc s_a^nds_0\wedge ds_b\wedge ds_c
\en
where $(a,b,c)=(1,2,3)$, $(2,3,1)$, $(3,1,2)$.
{}From this we see that
\be\label{KERN}
&&{\rm Ker}(\bar Z^2\buildrel d\over\rightarrow\bar Z^3)=
\sum_{0\leq \alpha<\beta<\gamma\leq3}\Ac 
s_\alpha s_\beta s_\gamma \omega
\oplus\sum_{m,n\geq1\atop m,n\not=2}
\sum_{0\leq \alpha<\beta \leq3}\Kc s_\alpha^m s_\beta^n \omega\\
&&\oplus\sum_{n\geq1\atop n\not=2}
\left(\sum_{1\leq a\not=b\leq3}\Kc(s_0^2+j_{ab}s_b^2)s_a^n\omega
+\sum_{1\leq a\leq2}\Kc(s_a^2-s_{a+1}^2)s_0^n\omega\right)\\
&&\oplus\sum_{1\leq a<b\leq3}\Kc(j_{ab}s_a^2s_b^2+s_0^2s_a^2-s_0^2s_b^2)\omega.
\en
We know that the $\Fc$-module $\bar Z^2/
{\rm Im}(\bar Z^1\buildrel d\over\rightarrow\bar Z^2)$ has the free generators
$\kappa_0=k_0s_0\omega$, $\kappa_a=(k_1-j_ak_0)s_a\omega$ $(a=1,2,3)$ and
$\rho_a=(k_1-j_ak_0)s_a^2\omega$ $(a=1,2,3)$. For each degree $e\geq3$
and color $c=0,1,2,3$,
we want to construct elements of degree $e$ and color $c$ in
${\rm Ker}(\bar Z^2\buildrel d\over\rightarrow\bar Z^3)$ by taking $\Kc$-linear
combinations of $k_0^mk_1^n\kappa_\alpha$ $(\alpha=0,1,2,3)$ and
$k_0^mk_1^n\rho_a$ $(a=1,2,3)$. 
We set $\rho_0=k_0s_0^2\omega$.
We have the relation $\rho_0=\rho_1+\rho_2+\rho_3$.
A straightforward calculation shows that for $e=3,4$ we have none;
for $e=5$ we find $\xi_0=k_0\kappa_0$ for color $0$
and $\xi_a=(k_1-j_ak_0)\kappa_a$ for color $a=1,2,3$;
for $e=6$ we find two color $0$ elements:
\be
\xi_4&=&(k_1-j_1k_0)\rho_1-(k_1-j_2k_0)\rho_2-j_{1,2}k_0\rho_0,\\
\xi_5&=&(k_1-j_2k_0)\rho_2-(k_1-j_3k_0)\rho_3-j_{2,3}k_0\rho_0.
\en
For $e=5,6$ the above elements span the degree $e$ cohomology classes.
For $e=8$, we have $4$ obvious elements
$k_i\xi_j$ $(i=0,1;j=4,5)$, and in addition, we find
\be
\xi_6=\frac14k_0^2\sum_{a=1}^3\sum_{b\not=a}j_{a,b}\rho_a
+k_0\sum_{a=1}^3(k_1-j_ak_0)\rho_a.
\en
These $5$ elements span the degree $8$ cohomology classes.

Finally, we show that $\xi_a$ $(0\leq a\leq6)$
are the generators of the $\Fc$-module
${\rm Ker}(\bar Z^2\buildrel d\over\rightarrow\bar Z^3)/
{\rm Im}(\bar Z^1\buildrel d\over\rightarrow\bar Z^2)$.
For odd $e=2n+3\geq7$, let us consider the color $0$ case.
The cases of other colors are similar.
The degree $2n+3$ and color $0$ space in
$\bar Z^2/{\rm Im}(\bar Z^1\buildrel d\over\rightarrow\bar Z^2)$
has the $\Kc$-basis consisting of $n+1$ elements
$k_0^jk_1^{(n-j)}\kappa_0$ $(0\leq j\leq n)$. We have
\be
d(k_1^n\kappa_0)=3s_0^{2n}ds_1\wedge ds_2\wedge ds_3\in\bar Z^3.
\en
Therefore, the $\Kc$-dimension of the degree $2n+3$ and 
color $\bar{0}$ subspace
of ${\rm Ker}(\bar Z^2\buildrel d\over\rightarrow\bar Z^3)/
{\rm Im}(\bar Z^1\buildrel d\over\rightarrow\bar Z^2)$ is $n$.
Since the $\Kc$-linearly independent elements
$k_0^jk_1^{n-1-j}\xi_0$ $(0\leq j\leq n-1)$ belongs to this subspace, they
span the subspace.

For even $e=2n+4\geq10$, the degree $2n+4$ subspace of
$\bar Z^2/{\rm Im}(\bar Z^1\buildrel d\over\rightarrow\bar Z^2)$
has the $\Kc$-basis consisting of $3(n+1)$ elements
$k_0^jk_1^{(n-j)}\rho_a$ $(0\leq j\leq n;a=1,2,3)$.
A simple calculation shows that the elements
$d(k_1^n\rho_0)$, $d(k_0^n\rho_a)$ $(a=1,2,3)$ are $\Kc$-linearly independent
in $\bar Z^3$. Therefore, the $\Kc$-dimension of the degree $2n+4$ subspace of
${\rm Ker}(\bar Z^2\buildrel d\over\rightarrow\bar Z^3)/
{\rm Im}(\bar Z^1\buildrel d\over\rightarrow\bar Z^2)$ is $3n-1$.
On the other hand, we obtain $3n-1$ independent elements of the subspace from
the degree $6$ elements $\rho_4,\rho_5$ and the degree $8$ element $\rho_6$.
We conclude that they span the subspace.
\end{proof}

We have explicitly constructed free bases for the spaces
$\Hc{}^{\scriptscriptstyle[0]}\supset
\Hc{}^{\scriptscriptstyle[-1]}\supset
\Hc{}^{\scriptscriptstyle[-2]}$.
In \cite{Sa4}, similar spaces are studied for
hypersurface singularities $\tilde E_6$, $\tilde E_7$, $\tilde E_8$.
The results in this appendix supplement some part of his construction
in $\tilde D_5$, where the elliptic singularity is given by
a complete intersection of two quadrics.

\section{Cycles and integrals in the classical limit}
\label{app:cycles}

In the classical limit, 
the defining relations of the Sklyanin algebra 
turns into \eqref{QUAD1}, \eqref{QUAD2}
which define an affine algebraic surface $\S\subset \C^4$.
In this section, we study the classical limit of 
the functional $\Tr_\la$ and identify it with 
an integral over a cycle on $\S$. 

Sklyanin's formulas for the representation \eqref{eq:rep(a)} 
gives an explicit uniformization of $\S$.   
Consider the limit 
$$
\eta\to 0,\quad \la\to\infty,\quad \eta\la \equiv 
\mu \ \text{finite}.  
$$
In the right hand side of \eqref{eq:rep(a)}, we  
replace $\eta\partial_u$ by $-2\pi i v$ where $v$ is the 
variable canonically conjugate to $u$, 
$$
2\pi \{v,u\}=1.   
$$
In the limit, the formulas for 
$s_0=S_0$, $s_a=\eta S_a$ ($a=1,2,3$) tend to 
\begin{align}
&s_{\al}(u,v)=c_{\al}\frac{\theta _{\al +1}(2u-\mu)e^{-2\pi i v}-
\theta _{\al +1}(-2u-\mu)e^{2\pi i v}}{\theta _1(2u)}, 
\label{clrep}
\end{align}
where $(u,v)\in \mathbb{C}^2$ and 
$$c_0=\frac 1 2, \quad
c_a=\sqrt{\varepsilon _a}
\ \ \frac {\theta _{a+1}(0)}{2\theta _1'}\ \text{ for}\  a=1,2,3. $$
Eq. \eqref{clrep} provides a parametrization of 
the surface \eqref{QUAD1},\eqref{QUAD2} with 
\begin{align}
k_0=\left(\frac{\theta _1(\mu)}{\theta _1'}\right)^2,\quad 
k_1=
\left(\frac{\theta _1(\mu)}{\theta _1'}\right)^2
\partial ^2\log\theta _1(\mu).
\label{cascl}
\end{align}

For definiteness, we consider the case $\tau\in i\R_{>0}$, $0<\mu<1/2$. 
The functions $s_{\alpha}$ have common periods 
$$ 
e_1=(1,0),\quad  e_2 = (0,1),\quad e_3=(\tau,2\mu),
$$
so $(u,v)$ should be regarded as variables on $\C^2/(\sum_{i=1}^3\Z e_i)$. 
We shall consider the fundamental domain:
$$ 
0\le \Re(u)<1,\quad 0\le \Im (u)<\frac  1 i \tau,
\quad 0\le \Re (v)<1. 
$$
There are also pole divisors of $s_\alpha$
at $u=p_i$, where 
\begin{align}
p_0=0,
~~p_1=\frac{1}{2},
~~
p_2=\frac{1+\tau}{2},
~~p_3=\frac{\tau}{2}.
\label{points}
\end{align}
If we neglect these divisors, 
there are, obviously, three non-trivial 2-cycles which are tori with
generators $(e_1,e_3)$, $(e_1,e_2)$, $(e_2,e_3)$.  
We denote them by $\gamma _0$, $\gamma _1$, $\gamma _2$. 
One can choose the tori $\gamma _1$, $\gamma _2$ 
so that they do not intersect with the pole divisors. 
As for $\gamma _0$, the first impression is that it hits the divisors. 
But actually one has to be very careful at this point. 
{}From the formulae (\ref{clrep}) it follows that 
there are no singularities at (\ref{points}) if 
$u=p_0,p_1$ and $v=0,1/2 $, 
or $u=p_2,p_3$ and $v=\mu,\mu+1/2$.
Hence the actual divisors are 
$\mathcal{D}=\mathcal{D}'\cup\mathcal{D}''$, with
\bea
&&
\mathcal{D}'=\cup_{i=0,1}\{(p_i,v)\mid 0\le \Re (v)<1,~~v\neq 0,1/2\},
\label{div}
\\
&&
\mathcal{D}''=\cup_{i=2,3}\{(p_i,v)\mid 0\le \Re (v)<1,~~v\neq \mu,\mu+1/2\}.
\nn
\ena
This means, first of all, that we can 
modify  $\gamma _0$ at $v=0$ and $v=\mu$ into a well-defined 
cycle without intersection with $\mathcal{D}$, 
as depicted in {\it fig. 1} below.

\vskip 1cm
\hskip 3cm  
\includegraphics[scale=.6]{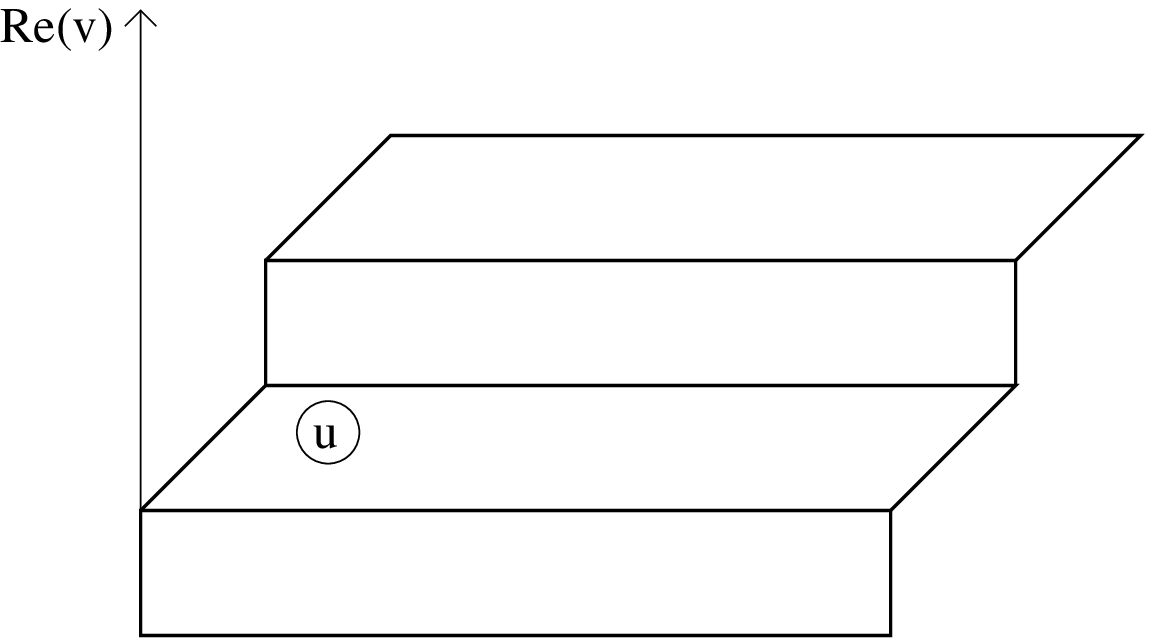}
\hfill{\it fig. 1}
\vskip 0.4cm
\noindent

Now we have another possibility. 
We can draw spheres $\delta _0,\delta _1$ which have 
as south (resp. north) poles  the points 
$(p_0,0)$, $(p_1, 0)$ (resp. $(p_0,1/2)$, $(p_1, 1/2)$).   
In the vicinity of these points they are parallel to the $u$-plane, 
and every section of it by the plane $\Re (v)=a$ for $0<a<1/2$ 
is a cycle around $p_0,p_1$ in the $u$-plane. 
\vskip 1cm
\hskip 3cm  
\includegraphics[scale=.7]{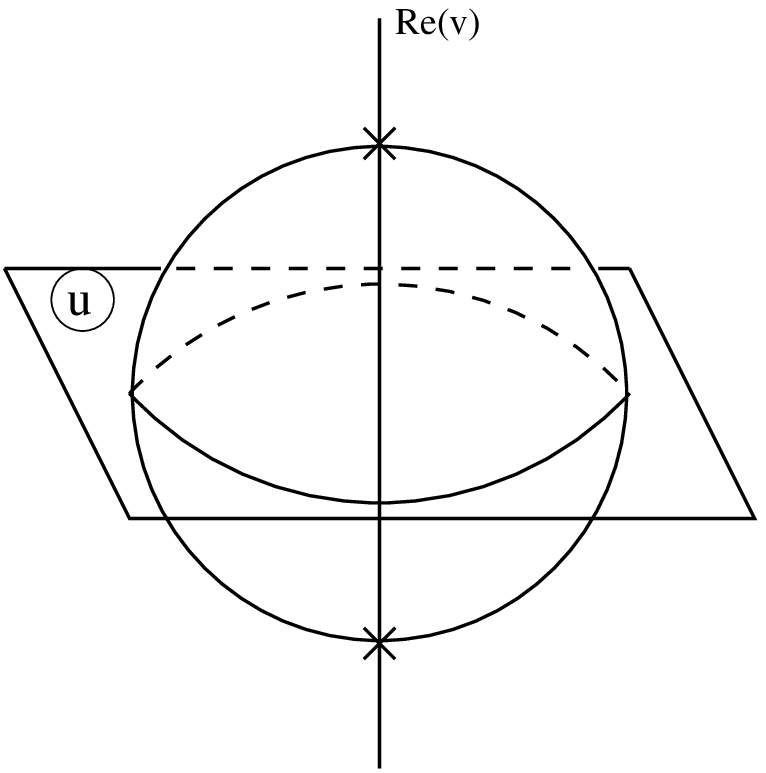}
\hfill{\it fig. 2}
\vskip 0.2cm
\noindent
These spheres do not intersect $\mathcal{D}$. 
Similarly, we construct spheres $\delta _2,\delta _3$ which have 
as south (resp. north) poles the points $(p_2,\mu)$, $(p_3, \mu)$ 
(resp. $(p_2,\mu +1/2)$, $(p_3, \mu+1/2)$).

The homology group $H_2(\S,\Z)\simeq\Z^7$ is known.
We will see below that our cycles
$\delta_0,\delta_1,\delta_2,\delta_3,\gamma_0,\gamma_1,\gamma_2\in H_2(\S,\Z)$
give a linearly independent basis in $H_2(\S,\C)$.

Using our seven monomials we construct 2-forms
\begin{align}
&\omega =-\frac{1}{4\pi} \frac {ds_1\wedge ds_2}{s_0s_3},\quad
\omega _0= s_0^2 \ \omega,\quad
\omega _3= s_3^2 \ \omega, 
\nn\\
&\sigma _{\alpha}=
 s_{\al} \ \omega,\quad \al =0,1,2,3.
\nn
\end{align}
By direct computation we check that 
in Sklyanin's parameterization (\ref{clrep})
\begin{align}
\omega=du\wedge dv. 
\label{meas} 
\end{align}
This formula allows us to calculate the integrals explicitly.  

Let us start with the cycles $\delta _{\al}$. We obtain 
\begin{align}
\int\limits _{\delta _{\alpha}}\omega =
\int\limits _{\delta _{\alpha}}\omega _{\beta}=0, 
\quad
\int\limits _{\delta _{\alpha}}\sigma _{\beta}=
2
\epsilon _{\al\beta}c_{\beta}\frac{\theta _{\beta +1}(-\mu )}{\theta '_1}, 
\nn
\end{align}
where $\epsilon _{\al\beta}$ are elements of the matrix:
$$
\epsilon =\begin{pmatrix} 1&\ \ 1&\ \ 1&\ \ 1\\1&\ \ 1&-1&-1\\1&-1&\ \ 1&-1
\\1&-1&-1&\ \ 1 \end{pmatrix}.
$$
Note, in particular, that
\begin{align}
\int\limits _{\sum_{\alpha=0}^3\delta _{\alpha}}\sigma _{\beta}=0
\end{align}
for $\beta=1,2,3$.

For the integrals over $\gamma_k$ $(k=1,2)$, we find
\begin{align}
&\int\limits _{\gamma _{k}}\sigma_{\alpha}=0, \quad
\int\limits _{\gamma _1 }\omega =1,\quad
\int\limits _{\gamma _2}\omega =\tau, 
\nn\\
&\int\limits _{\gamma _{1}}\omega _{\alpha} =
-2\left(\frac{c_{\alpha}\theta _{\al +1}(\mu)}{\theta _1'}\right)^2
\partial ^2\log\theta _{\al +1}(\mu),
\nn
\\
&\int\limits _{\gamma _{2}}\omega _{\alpha} =-2
\left(\frac{c_{\alpha}\theta _{\al +1}(\mu)}{\theta _1'}\right)^2
\(\tau\partial ^2\log\theta _{\al +1}(\mu)+2\pi i\),
\nn
\end{align}
For $\gamma_0$, we introduce
$$
\tilde{\gamma}_0=\gamma _0-2\mu\gamma _1+\frac 1 2 \sum\limits _{\beta =0}^3
\delta _{\beta}.
$$
Then, we have
\begin{align}
&\int\limits _{\tilde{\gamma}_0}\omega
=\int\limits _{\tilde{\gamma}_0}\sigma _{\al}=0,\quad
\int\limits _{\tilde{\gamma}_0}\omega _{\al}=
2 \left(\frac{c_{\alpha}\theta _{\al +1}(\mu)}{\theta _1'}\right)^2
\partial\log\theta _{\al +1}^2(\mu),
\nn
\end{align}
Using the above formulae we can calculate the determinant of the period-matrix as 
\begin{align}
\det \(\mathcal {P}\)
=\text{Const}\cdot \theta _1(2\mu)\(\theta _1(\mu)\theta _0(\mu)\)^2
\frac{\partial}{\partial \mu}\log 
\(\frac {\theta _1(\mu)}{\theta _0(\mu)}\),\nn
\end{align}
whence we conclude that for generic $\mu$ 
the period matrix is non-degenerate. 

Comparison with the quantum formulae \eqref{eq:trSa1}, \eqref{eq:trSa2}
shows the exceptional role of the cycle $\gamma _0$ as $\eta\to0$, 
\begin{align}
&\text{Tr} _{\la}(S_\alpha)\sim 
\ \ \frac {1}{2 \eta}
\int\limits _{\gamma _0}s_\alpha\ 
\omega
\nn\\
&\text{Tr} _{\la}(S_{\alpha}^{2})\sim 
\ \ 
\frac 1{2 \eta}
\int\limits _{\gamma _0}s_\alpha^2\ 
\omega 
\times 
\begin{cases}
1& (\alpha=0),\\
\frac 1 {\eta ^2}& (\alpha=1,2,3).\\
\end{cases}
\nn
\end{align} 
Actually, classical limits of 
all the elements of 
$F$ 
in \eqref{eq:fa1}
--\eqref{eq:fa3} 
can be found among integrals over the cycles $\gamma _0,\gamma _1,\gamma _2$.
This fact played an important heuristic role in the 
calculation of traces.
\bigskip

\section{Technical Lemmas}\label{app:lemmas}

In this Appendix, we collect some technical 
matters related to $\Tr_\la$. 

The first is the color conservation for $\Tr_\la$. 
\begin{lem}\label{lem:color}
If $A\in\A^{(m,n)}$, then 
\be
\Tr_\la A=0\qquad \mbox{unless $(m,n)=(0,0)$}.
\en
\end{lem}
\begin{proof}
Set 
\be
\varphi_1(f)(u):=f(u+\frac{\tau}{2})e^{2\pi i k u},
\qquad
\varphi_2(f)(u):=f(u+\frac{1}{2}).
\en
One verifies easily that 
$\varphi_1,\varphi_2\in\End(\V^{(k)})$ and 
\be
&&\varphi_1\circ\pi^{(k)}(S_\alpha)\circ\varphi_1^{-1}
=(-1)^{\bar{\alpha}_1}\pi^{(k)}(S_\alpha),
\\
&&\varphi_2\circ\pi^{(k)}(S_\alpha)\circ\varphi_2^{-1}
=(-1)^{\bar{\alpha}_2}\pi^{(k)}(S_\alpha),
\en
where $\bar{\alpha}=(\bar{\alpha}_1,\bar{\alpha}_2)\in\Z_2\times\Z_2$. 
It follows that 
\be
\tr_{V^{(k)}}\pi^{(k)}(A)
=(-1)^m\tr_{V^{(k)}}\pi^{(k)}(A)
=(-1)^n\tr_{V^{(k)}}\pi^{(k)}(A), 
\en
whence the lemma. 
\end{proof}

The next Lemma is concerned about the uniqueness of the 
representation of $\Tr_\la$. 

\begin{lem}\label{lem:unique}
Assume $\Im\eta, \Im\tau>0$, $\eta\not\in\Q+\Q\tau$. 
Let $g_i$ $(i=1,2,3)$ be elliptic functions with periods $1$ and $\tau$, 
and set $f(u)=\zeta(u)g_1(u)+u g_2(u)+g_3(u)$, 
where $\zeta(u)=-(1/2\pi i)\theta_1'(u)/\theta_1(u)$. 
If there exists an $N>0$ such that 
$f(k\eta)=0$ holds for all integers $k>N$, 
then we have $g_i(u)\equiv 0$ $(i=1,2,3)$. 
\end{lem}
\begin{proof}
We divide into three cases, (i)$g_1(u)=g_2(u)=0$,
(ii)$g_1(u)\not\equiv 0$ and $g_2(u)=0$, (iii) $g_2(u)\not \equiv 0$. 

Since $K=\{k\eta\mid k>N\}$ has accumulation points,
the assertion is evident in case (i). 
Let us show that (ii), (iii) lead to contradictions. 

In case (ii), considering $f(u)/g_1(u)$ we may assume $g_1(u)=1$. 
Then $f(u+\tau)=f(u)+1$. 
Choose a point $u_0\in\C\backslash(K\cup L)$ 
which is not a pole of $f(u)$. One can find a sequence of integers
$k_1<k_2<\cdots$, 
$k_n\to\infty$, such that $k_n\eta$ tends to $u_0$ in $\C/L$. 
Since $\Im\eta>0$, if we write $k_n\eta=a_n+b_n\tau$ ($a_n,b_n\in\R$), then 
the integer part of $b_n$ diverges. Therefore $f(k_n\eta)$ diverges as 
$n\to\infty$, which contradicts to the assumption $f(k_n\eta)=0$. 

In case (iii), we may assume $g_2(u)=1$. 
Set $F(u)=f(u+\eta)-f(u)$. We have $F(k\eta)=0$ ($k>N$), 
$F(u+1)=F(u)$, and $F(u+\tau)-F(u)=g_1(u+\eta)-g_1(u)$. 
Hence $F(u)=G(u)+\zeta(u)(g_1(u+\eta)-g_1(u))$ with some elliptic function
$G(u)$. From cases (i), (ii) we conclude that $F(u)=0$. 
In particular $g_1$ is a constant. Considering $f'(u+\eta)=f'(u)$, 
we find that $f(u)$ is a linear function. Clearly this is impossible. 
\end{proof}

Let us sketch the derivation of the formulas 
for $\Tr_\la S_\alpha$, $\Tr_\la S_\alpha^2$.  
We have the standard functional relation 
\be
&&t^{(1)}\Bigl(t-\frac{k}{2}\eta\Bigr)
t^{(k)}\Bigl(t+\frac{1}{2}\eta\Bigr)
=
\phi(t+\eta)t^{(k+1)}(t)
+\phi(t)t^{(k-1)}(t+\eta)
\label{eq:func-rel}
\en
for the transfer matrices
\be
&&t^{(k)}(t):=
\tr_{\V^{(k)}}
\left(r^{(k,1)}_{a,N}(t-t_N)\cdots r^{(k,1)}_{a,1}(t-t_1)\right)
\quad \in\End(V^{\otimes N}), 
\en
where 
$r^{(k,1)}(t):=(\pi^{(k)}\otimes\id)L(t)$ and
$\phi(t)=\prod_{j=1}^N[t-t_j-(k/2)\eta]$.  
Choosing $N=1,t_1=0$ and applying \eqref{eq:func-rel},  
we easily find \eqref{eq:trSa1}.  
The same method is applicable for \eqref{eq:trSa2}. 
We find it slightly simpler to use the difference equation 
for the matrices $\Xh^{(1,2)}_{a,2}$. 

\section{Transformation properties of $\Xh^{(i,j)}_n$}
\label{app:transX}

Let us study the transformation properties of $\Xh^{(i,j)}_n$
with respect to the shift of variables by half periods. 
For that purpose we exploit the order 4 
automorphisms $\iota^1,\iota^3$ of the Sklyanin algebra $\A$ 
given by  
\be
&&
\iota^1(S_0)=-\frac{\theta_1(\eta)}{\theta_2(\eta)}S_1,
~
\iota^1(S_1)=\frac{\theta_2(\eta)}{\theta_1(\eta)}S_0,
~
\iota^1(S_2)=i\frac{\theta_3(\eta)}{\theta_0(\eta)}S_3,
~
\iota^1(S_3)=-i\frac{\theta_0(\eta)}{\theta_3(\eta)}S_2,
\label{eq:auto1}
\\
&&
\iota^3(S_0)=\frac{\theta_1(\eta)}{\theta_0(\eta)}S_3,
~
\iota^3(S_1)=\frac{\theta_2(\eta)}{\theta_3(\eta)}S_2,
~
\iota^3(S_2)=-\frac{\theta_3(\eta)}{\theta_2(\eta)}S_1,
~
\iota^3(S_3)=\frac{\theta_0(\eta)}{\theta_1(\eta)}S_0.
\label{eq:auto2}
\en
In terms of the $L$-operator they can be written as 
\bea
&&\iota^1\left(L(t)\right)=L\Bigl(t+\frac{1}{4}\Bigr)\sigma^1,
\label{eq:Liota1}\\
&&\iota^3\left(L(t)\right)
=L\Bigl(t+\frac{\tau}{4}\Bigr)\sigma^3
 \times (-i)e^{\pi i (2t+\eta+\tau/4)}.
\label{eq:Liota3}
\ena

\begin{lem}\label{lem:trans-trA}
Let $A\in\A_n$ be an element of the Sklyanin algebra of even 
degree $n$, and let 
$\theta_1(t)^{-n}
\Tr_{t/\eta} A=g_{A,1}(t)-(t/\eta) g_{A,2}(t)$ 
where $g_{A,1},g_{A,2}$ are as in \eqref{eq:trdec}.
Then
\be
&&
\theta_1(t)^{-n}
\Tr_{\frac{t}{\eta}+\frac{1}{2\eta}}\iota^1(A)=g_{A,1}(t)
-\left(\frac{t}{\eta}+\frac{1}{2\eta}\right)g_{A,2}(t),
\\
&&
\theta_1(t)^{-n}
\Tr_{\frac{t}{\eta}+\frac{\tau}{2\eta}}\iota^3(A)
\\
&&\quad=
\left(
g_{A,1}(t)+
\frac{1}{2}g_{A,3}(t)
-\left(\frac{t}{\eta}+\frac{\tau}{2\eta}\right)g_{A,2}(t)\right)
\times\Bigl(-e^{-\pi i(\tau/2+2t)}\Bigr)^{n/2},
\\
\en
where 
$g_{A,3}(t)=g_{A,1}(t+\tau)-g_{A,1}(t)$
is an elliptic function.  
\end{lem}
\begin{proof}
In view of Theorem \ref{SEVEN}, it is enough to 
consider elements $A$ of the form $m\cdot S_\alpha^2$, 
where $m$ is a polynomial in $K_0,K_2$ of degree $(n-2)/2$. 
For $n=2$, the assertion can be verified 
{}from the explicit formula \eqref{eq:trSa2}.

Let $I_t$ denote the two-sided ideal of $\A$ generated by 
$K_0-4\theta_1(t)^2/\theta_1(2\eta)^2$,
$K_2-4\theta_1(t+\eta)\theta_1(t-\eta)
/\theta_1(2\eta)^2$, and let $\varpi_t:\A\to\A/I_t$ be the projection. 
{}From \eqref{eq:LCas} and \eqref{eq:Casval} 
we have
\be
\varpi_t\left(
L_1\Bigl(\frac{s}{2}\Bigr)L_2\Bigl(\frac{s}{2}-\eta\Bigr)
\right)\cP^-_{12}
=-\frac{\theta_1(t-s)\theta_1(t+s)}
{\theta_1^2(2\eta)}\cP^-_{12}.
\en
Along with \eqref{eq:Liota1} and \eqref{eq:Liota3} 
it follows that for $i=0,2$ 
\be
&&\varpi_{t+1/2}\Bigl(\iota^1(K_i)\Bigr)
=\varpi_{t}(K_i), \\
&&\varpi_{t+\tau/2}\Bigl(\iota^3(K_i)\Bigr)
=\varpi_{t}(K_i)\times
(-1)e^{-\pi i(\tau/2 +2t)}.
\en
The Lemma follows from these relations. 
\end{proof}

\begin{prop}
The $\Xh^{(i,j)}_{a,n}$ 
obey the following transformation laws.
\bea
&&\label{eq:transX1}
\\
&&
\sigma_{\bar k}^1\sigma_k^1
\Xh_{a,n}^{(i,j)}(\cdots,t_k+\frac{1}{2},\cdots)
\nn\\
&&\quad=
\Xh_{a,n}^{(i,j)}(\cdots,t_k,\cdots)
\times
\begin{cases}
\sigma_k^1\sigma_{\bar k}^1 & (k\neq i,j),\\
(-1)^{n-1}\prod_{p(\neq i,j)}\sigma_p^1\sigma_{\bar p}^1& (k=i,j),\\
\end{cases}
\nn
\\
&&\label{eq:transX2}
\\
&&
\sigma_{\bar k}^3\sigma_k^3
\Xh_{a,n}^{(i,j)}(\cdots,t_k+\frac{\tau}{2},\cdots)
\nn\\
&&\quad
=
\begin{cases}
\Xh_{a,n}^{(i,j)}(\cdots,t_k,\cdots)\sigma_k^3\sigma_{\bar k}^3
& (k\neq i,j),\\
\left(\Xh_{a,n}^{(i,j)}(\cdots,t_k,\cdots)
\pm\frac{1}{2}\delta_{a1}\Xh_{3,n}^{(i,j)}(\cdots,t_k,\cdots)
\right)
\times (-1)^{n-1}\prod_{p(\neq i,j)}\sigma_{\bar p}^3\sigma_p^3,
&
(k=i,j).\\
\end{cases}
\nn
\ena
In the last line the upper (resp. lower) sign in chosen for $k=i$
(resp. $k=j$). 
In particular, $\Xh_{2,n}^{(i,j)},
\Xh_{3,n}^{(i,j)}$ are elliptic functions 
of $t_1,\cdots,t_n$ with periods
$1$, $\tau$. 
\end{prop}
\begin{proof}
It is enough to prove the case $(i,j)=(1,2)$. 
If $k\neq 1,2$, this is a simple consequence of the 
transformation law of the $L$-operator
\be
&&
L\left(t+\frac{1}{2}\right)
=-\sigma^1 L(t)\sigma^1,
\\
&&
L\left(t+\frac{\tau}{2}\right)
=-\sigma^3 L(t)\sigma^3\times
e^{-2\pi i(2t+\eta+\tau/2)}.
\en

Consider the case $k=1$. Using the automorphism $\iota^1$
we have 
\be
&&\Xh^{(1,2)}_n(t_1+\frac{1}{2},\cdots)\times
[t_{12}+1/2]\prod_{p=3}^n
[t_{1p}+1/2][t_{2p}]
\\
&&
=\Tr_{\frac{t_{12}}{\eta}+\frac{1}{2\eta}}
\left(
T^{[1]}\Bigl(\frac{t_1+t_2}{2}+\frac{1}{4};
t_1+\frac{1}{2},\cdots,t_n\Bigr)
\right)
P_{12}\cP^-_{1\bar1}\cP^-_{2\bar2}
\\
&&=
\Tr_{\frac{t_{12}}{\eta}+\frac{1}{2\eta}}
\iota^1\left(
T^{[1]}\Bigl(\frac{t_1+t_2}{2};t_1,\cdots,t_n\Bigr)
\right)\prod_{p=2}^n\sigma_{\bar p}^1\sigma_p^1\,
P_{12}\cP^-_{1\bar1}\cP^-_{2\bar2}
\\
&&
=
\sigma_{\bar1}^1\sigma_1^1
\Tr_{\frac{t_{12}}{\eta}+\frac{1}{2\eta}}
\iota^1\left(
T^{[1]}\Bigl(\frac{t_1+t_2}{2};t_1,\cdots,t_n\Bigr)
\right)
P_{12}\cP^-_{1\bar1}\cP^-_{2\bar2}
\prod_{p=3}^n\sigma_{\bar p}^1\sigma_p^1.
\en
Applying Lemma \ref{lem:trans-trA} we obtain 
\eqref{eq:transX1} with $k=1$.  
Eq. \eqref{eq:transX2} is shown similarly. 
The case $k=2$ can be obtained by using the translation 
invariance. 
\end{proof}
\bigskip

\noindent
{\it Acknowledgments.}\quad
Research of HB is supported 
by the RFFI grant \#04-01-00352.
Research of MJ is 
supported by 
the Grant-in-Aid for Scientific Research B2--16340033
and 
A2--14204012.
Research of TM is 
supported by 
the Grant-in-Aid for Scientific Research A1--13304010.
Research of 
FS is supported by INTAS grant \#03-51-3350 and by 
EC networks  "EUCLID",
contract number HPRN-CT-2002-00325 and "ENIGMA",
contract number MRTN-CT-2004-5652. 
Research of YT is 
supported by University of 
Tsukuba Research Project. 
This work was also supported by the grant of 21st Century 
COE Program at Graduate School of Mathematical Sciences, 
the University of Tokyo,
and at RIMS, Kyoto University. 

MJ, TM and YT are grateful to 
K. Saito for explaining his theory of simple-elliptic 
singularities. 
MJ thanks 
V. Mangazeev, A. Zabrodin, A. Mironov, C. Korff 
and R. Weston for discussions. 
HB thanks M. Lashkevich for tight cooperation in numerical 
checking. 
He also thanks F. G{\"o}hmann, A. Kl{\"u}mper 
and K. Fabricius for discussions.
Last but not least, 
HB and FS wish to thank the University of Tokyo
for the invitation and hospitality 
where the present work was started, 
and HB, JM, TM and YT to Universit{\'e} Paris VI for hospitality, 
where it was completed.

\end{document}